\documentclass[12pt]{article}

\pdfoutput=1

\usepackage{cite}

\usepackage{amssymb,latexsym}

\usepackage{epstopdf}

\usepackage{graphics,epsfig}

\usepackage{color}

\usepackage{mathrsfs}

\DeclareMathAlphabet{\mathpzc}{OT1}{pzc}{m}{it}

\let\a=\alpha \let\b=\beta \let\g=\gamma \let\d=\delta \let\e=\epsilon
\let\z=\zeta  \let\th=\theta  \let\k=\kappa
\let\l=\lambda \let\m=\mu \let\n=\nu \let\x=\xi \let\p=\pi 
\let\s=\sigma   \let\f=\phi  
 
      \let\G=\Gamma \let\D=\Delta \let\Th=\Theta 
\let\X=\Xi  \let\S=\Sigma  \let\Y=\Psi
 
\let\la=\label  
  
\def\nn{\nonumber} \def\bd{\begin{document}} \def\ed{\end{document}}
\def\ds{\documentstyle} \let\fr=\frac \let\bl=\bigl \let\br=\bigr
\let\Br=\Bigr \let\Bl=\Bigl
\let\bm=\bibitem
\let\na=\nabla
\def\tU{{\widetilde U}}
\let\pa=\partial \let\ov=\overline
\def\ie{{\it i.e.\ }}
\newcommand{\be}{\begin{equation}}
\newcommand{\ee}{\end{equation}}
\def\ba{\begin{array}}
\def\ea{\end{array}}
\def\ft#1#2{{\textstyle{{\scriptstyle #1}\over {\scriptstyle #2}}}}
\def\fft#1#2{{#1 \over #2}}
\def\F#1#2{{ F_{#1}^{(#2)} }}
\def\cF#1#2{{ {\cal F}_{#1}^{(#2)} }}

\def\R{{\bf R}}
\def\sst#1{{\scriptscriptstyle #1}}
\def\oneone{\rlap 1\mkern4mu{\rm l}}
\def\e7{E_{7(+7)}}
\def\td{\tilde}
\def\wtd{\widetilde}
\def\im{{\rm i}}
\def\bog{Bogomol'nyi\ }
\newcommand{\ho}[1]{$\, ^{#1}$}
\newcommand{\hoch}[1]{$\, ^{#1}$}
\newcommand{\bea}{\begin{eqnarray}}
\newcommand{\eea}{\end{eqnarray}}
\newcommand{\ra}{\rightarrow}
\newcommand{\lra}{\longrightarrow}
\newcommand{\Lra}{\Leftrightarrow}
\newcommand{\ap}{\alpha^\prime}
\newcommand{\bp}{\tilde \beta^\prime}
\newcommand{\cB}{{\cal B}}
\newcommand{\cO}{{\cal O}}
\newcommand{\vecx}{\vec{x}}
\newcommand{\vecy}{\vec{y}}
\newcommand{\vecp}{\vec{p}}
\newcommand{\vecq}{\vec{q}}
\newcommand{\tr}{{\rm tr} }
\newcommand{\Tr}{{\rm Tr} }
\newcommand{\NP}{Nucl. Phys. }

\newcommand{\cL}{{\cal L}}
\newcommand{\cA}{{\cal A}}
\newcommand{\cT}{{\cal T}}
\newcommand{\cD}{{\cal D}}
\newcommand{\cH}{{\cal H}}

\def\sst#1{{\scriptscriptstyle #1}}
\def\0{{\sst{(0)}}}
\def\1{{\sst{(1)}}}
\def\2{{\sst{(2)}}}
\def\3{{\sst{(3)}}}
\def\4{{\sst{(4)}}}
\def\5{{\sst{(5)}}}
\def\6{{\sst{(6)}}}
\def\7{{\sst{(7)}}}
\def\8{{\sst{(8)}}}
\def\9{{\sst{(9)}}}
\def\p{{\sst{(p)}}}
\def\q{{\sst{(q)}}}
\def\ve{\varepsilon}
\def\vf{\varphi}
\def\F{\Phi}
\def\wg{\wedge}

\def\thb{\bar{\theta}}
\def\Thb{\bar{\Theta}}
\def\barp{\bar{p}}
\def\barq{\bar{q}}
\def\barc{\bar{c}}
\def\bard{\bar{d}}
\def\e{\epsilon}

\def \bi{\bibitem}
\def \la {\label}

\def \l {\lambda}
\def\foot{\footnote}
\def \tl  {{\tilde \l}}
\def \sql {{\sqrt \l}}
\def \adss {$AdS_5 \times S^5$\ }
\newcommand{\rf}[1]{(\ref{#1})}
\def \ov {\over}

\def\th{\theta}
\def\Th{\Theta}
\def\vth{\vartheta}
\def\btheta{{\bar\theta}}
\def\ttheta{{{\tilde\theta}}}
\def\bttheta{{{\bar\ttheta}}}
\def\vth{\vartheta}

\def\ra{\rightarrow}
\def\N{\nabla}
\def\F{{\cal F}}
\def\uM{\underline{M}}
\def\uA{\underline{A}}
\def\uN{\underline{N}}
\def\uP{\underline{P}}
\def\ua{\underline{a}}
\def\ub{\underline{b}}
\def\uc{\underline{c}}
\def\ud{\underline{d}}
\def\ue{\underline{e}}
\def\uf{\underline{f}}
\def\ui{\underline{i}}
\def\uj{\underline{j}}
\def\uk{\underline{k}}
\def\ul{\underline{l}}
\def\ual{\underline{\alpha}}
\def\ube{\underline{\beta}}
\def\um{\underline{m}}
\def\un{\underline{n}}
\def\up{\underline{p}}
\def\uq{\underline{q}}
\def\ur{\underline{r}}
\def\us{\underline{s}}
\def\umu{\underline{\mu}}
\def\unu{\underline{\nu}}
\def\ula{\underline{\l}}
\def\uka{\underline{\k}}
\def\usi{\underline{\s}}
\def\urh{\underline{\r}}
\def\cc{\circ}
\def\eqv{\equiv}

\def\ni{\noindent}

\def\Ep{E^{{}^{(+)}}}
\def\Em{E^{{}^{(-)}}}

\def\Mp{M^{{}^{(+)}}}
\def\Mm{M^{{}^{(-)}}}

\def \ha{{1\ov 2}}

\def\r{\rho}

\def\Y{{\rm Y}}
\def\X{{\rm X}}
\def\tY{\tilde{\rm Y}}
\def\tX{\tilde{\rm X}}
\def\dY{\dot{\rm Y}}
\def\dX{\dot{\rm X}}

\def \J {\mathcal{J}}
\def \del {\partial}

\def\dF{\dot{F}}
\def\dG{\dot{G}}
\def\df{\dot{f}}
\def \E {{\cal E}}
\def \S {{\cal S}}
\def \J {{\cal J}}

\def\ms{\mathcal{S}}
\def\mj{\mathcal{J}}
\def\soj{\fr{\ms}{\mj}}
\def \R {{\bf R}}
\def \om {\omega}
\def \bE {\bar E}
\def \x {{\cal X}}

\def \bi{\bibitem}
\def \la {\label}

\def \l {\lambda}
\def\foot{\footnote}
\def \tl  {{\tilde \l}}
\def \sql {{\sqrt \l}}
\def \adss {$AdS_5 \times S^5$\ }
\def \ov {\over}

\def \varpi {{\rm w}}

\def\thb{\bar{\theta}}
\def\Thb{\bar{\Theta}}
\def\mb{\bar{\m}}
\def\ab{\bar{\a}}
\def\zb{\bar{z}}
\def\psib{\bar{\psi}}
\def\barp{\bar{p}}
\def\barq{\bar{q}}
\def\barc{\bar{c}}
\def\bard{\bar{d}}
\def\e{\epsilon}
\def\wb{\bar{w}}
\def\lb{\bar{\l}}
\def\Jb{\bar{J}}
\def\Nb{\bar{N}}
\def\Zb{\bar{Z}}
\def\pab{\bar{\pa}}

\def\Cb{\bar{C}}

\def\At{\tilde{A}}
\def\Bt{\tilde{B}}
\def\Ct{\tilde{C}}
\def\Dt{\tilde{D}}
\def\Et{\tilde{E}}
\def\Ft{\tilde{F}}
\def\Gt{\tilde{G}}
\def\Ht{\tilde{H}}
\def\Kt{\tilde{K}}
\def\Mt{\tilde{M}}
\def\Nt{\tilde{N}}
\def\Rt{\tilde{R}}
\def\at{\tilde{a}}
\def\bt{\tilde{b}}
\def\ct{\tilde{c}}
\def\dt{\tilde{d}}
\def\et{\tilde{e}}
\def\ft{\tilde{f}}
\def\htil{\tilde{h}}
\def\gt{\tilde{g}}
\def\nt{\tilde{n}}
\def\mut{\tilde{\mu}}
\def\nut{\tilde{\nu}}
\def\pht{\tilde{\f}}
\def\vft{\tilde{\vf}}

\def\rht{\tilde{\rho}}

\def\asth{\hat{*}}
\def\phh{\hat{\phi}}

\def\bA{{\bf A}}

\def\ola{\overleftarrow}
\def\ora{\overrightarrow}
\def\alt{\tilde{\a}}

\def\eh{\hat{e}}
\def\eph{\hat{\e}}
\def\ph{\hat{p}}
\def\alh{\hat{\a}}
\def\beh{\hat{\b}}
\def\gah{\hat{\g}}
\def\Fh{\hat{F}}
\def\muh{\hat{\m}}
\def\nuh{\hat{\n}}
\def\thh{\hat{\th}}
\def\rhh{\hat{\r}}
\def\dh{\hat{d}}
\def\ih{\hat{i}}
\def\jh{\hat{j}}
\def\hh{\hat{h}}
\def\nh{\hat{n}}
\def\gh{\hat{g}}
\def\kh{\hat{k}}
\def\deh{\hat{\d}}
\def\wh{\hat{w}}
\def\lah{\hat{\l}}
\def\Ah{\hat{A}}
\def\Kh{\hat{K}}
\def\Nh{\hat{N}}
\def\Rh{\hat{R}}
\def\Ch{\hat{C}}
\def\Omh{\hat{\Omega}}

\def\xh{\hat{x}}

\def\ps{\rlap{\, /}\;\,p }
\def\ks{\rlap{\, /}\;\,k }

\def\gym{g_{YM}}

\def\adot{\dot{a}}
\def\bdot{\dot{b}}
\def\bpa{\bar{\pa}}

\def\pr{\prime}
\def\ssk{\medskip}
\def\clb{\color{blue}}
\def\clr{\color{red}}
\def\clg{\color{green}}

\begin{document}

\overfullrule=0pt
\parskip=2pt
\parindent=12pt
\headheight=0in \headsep=0in \topmargin=0in
\oddsidemargin=0in

\vspace{ -3cm}
\thispagestyle{empty}

 \vspace{0.1cm}

\setcounter{equation}{0}
\setcounter{footnote}{0}
\setcounter{section}{0}

\begin{center}

{\Large\bf  Lagrangian constraints and renormalization of 4D gravity}

\vskip 0.8cm

 \vspace{.5cm}

\vspace{0.5cm}
I. Y. Park
\\

\vspace{0.3cm}


{\it Department of Physics, Hanyang University \\
Seoul 133-791, Korea}\\

\vspace{0.3cm}
{\it Department of Applied Mathematics,
Philander Smith College 
                               \\
Little Rock, AR 72223, USA \\
inyongpark05@gmail.com
}

\end{center}

 \vspace{0.1cm}

\begin{abstract}
\ni It has been proposed in \cite{Park:2014tia} that 4D Einstein gravity 
becomes effectively reduced to 3D after solving the Lagrangian analogues of the Hamiltonian and momentum constraints of the Hamiltonian quantization. The analysis in \cite{Park:2014tia} was carried out at the classical/operator level. We review the proposal and make a transition to the path integral account. We then set the stage for explicitly carrying out the two-loop renormalization procedure of the resulting 3D action. We also address a potentially subtle issue in the gravity context concerning whether renormalizability does not depend on the background around which the original action is expanded.  
 
\end{abstract}

\newpage
\section{Introduction}
Rough horizon ideas \cite{Braunstein:2009my}\cite{Almheiri:2012rt} may eventually lead us to fill in some of the gaps and potentially bring more radical changes in our understanding of black hole physics. 
It was proposed in \cite{Almheiri:2012rt} that not all of the postulates of the black hole complementarity \cite{Susskind:1993if} may be mutually consistent. The arguments that oppose (e.g., \cite{Nomura:2011rb,Chowdhury:2012vd,Giddings:2012gc,Page:2013mqa,Verlinde:2013uja,Papadodimas:2013jku,Mathur:2012jk}) or support (e.g., \cite{Bousso:2012as,Avery:2013exa,Bousso:2013uka,Park:2014mba}) this so-called Firewall have mostly been indirect. See, e.g., \cite{Hwang:2012nn,VanRaamsdonk:2013sza,Kim:2013caa} for other related discussions as well as \cite{Kawai:2013mda,Almheiri:2013wka,Silverstein:2014yza} for the works that more directly address Firewall by using specific models.
We believe that it would have been possible to address Firewall much more directly if it had been known how to quantize gravity in a renormalizable way. Thus motivated by better understanding of Firewall, we have revisited quantization of 4D Einstein gravity in \cite{Park:2014tia}\cite{Park:2014qoa}.

Quantization of 4D gravity has a long history (\cite{DeWitt:1967ub,Kuchar:1993ne,Carlip:2001wq,Isham,Arnowitt:1962hi,Sen:1982qb,Ashtekar:1986yd,
Rovelli:1997yv,Thiemann:2007zz,Woodard:2014jba} and refs therein).
Past approaches can be grouped into two categories: ``canonical" and ``covariant." The canonical approach splits the spacetime into time and space, and historically Dirac's method was used. The covariant approach treats the time and space on an equal footing, thus relatively more covariant compared with the canonical approach, but does consider the theory around a fixed background. The present approach has elements from both the canonical and covariant approaches: it uses 3+1 splitting of the spacetime and starts with the ADM Hamiltonian quantization. However, instead of remaining entirely in the Hamiltonian quantization followed by the Wheeler-DeWitt equation, the present method employs the Lagrangian method.
 
Six metric components are kept in the conventional covariant attempts of quantizing gravity \cite{'tHooft:1974bx,Deser:1974cy,Deser:1974cz,Goroff:1985th,Anselmi:2003xb} wherein four out of ten components are gauge-fixed by the bulk diffeomorphism. The advantage of keeping six components, four of which are unphysical, lies in maintenance of the 4D covariance. It has recently been proposed in \cite{Park:2014tia}\cite{Park:2014qoa} that all of the eight unphysical components can explicitly be removed in the ADM formulation by exploiting the non-dynamical nature of the lapse function and shift vector. Gauge-fixing of these non-dynamical fields introduces constraints; 4D Einstein gravity reduces to 3D as a result of solving these constraints, and thereby a possibility for renormalizability of 4D Einstein gravity opens.

The proposal was based on the observation that there exists an elaborate gauge-fixing procedure that induces an effective reduction of the 4D Einstein-Hilbert action to the 3D Einstein-Hilbert action of a hypersurface. Unlike the constraints in the conventional Hamiltonian quantization, the shift vector and lapse function constraints admit an explicit solution in the Lagrangian formulation, an outcome that was crucial for the reduction that opens up the possibility of the renormalizability.  
The goal of the present work is to give a more detailed and refined account of \cite{Park:2014tia}, and to set the stage for an explicit two-loop renormalization procedure in the 3D description.

There seem to be several reasons why explicit removal of all of the unphysical degrees of freedom was not sought in the context of quantizing general relativity in the past even though such removal was considered in other contexts such as the radiation gauge used in the linearized gravity \cite{Smarr:1978dia}\cite{Wald}. 
Firstly, it was the success of renormalization of gauge theories in covariant gauges. Unlike gravity, gauge theories were shown to be renormalizable even in the presence of the unphysical fields running in the loops: the advantages offered by covariance outweighed the ``inefficiency" of keeping the non-dynamical fields. 
Even without the success of the gauge theories, the covariance would have been viewed as a device that helped one to recognize possible forms of the counterterms. Secondly, emphasis was somehow placed on the Hamiltonian formalism, while little attention was paid, as far as we are aware, to the implications of the Hamiltonian result for the Lagrangian formalism.  
Once incorporated into the Lagrangian setup, the result of the Hamiltonian analysis reveals, as we will review, that the physical degrees of freedom of the system are reduced to those of a hypersurface \cite{Park:2014tia}.

The search for the true degrees of freedom for gravity has a long history (see, e.g., \cite{York:1972pr,York:1972sj,Isham:1984rz,Engle:2009ba,Gerhardt:2012ku}).
The fact that the true degrees of freedom are those of a 3D hypersurface appeared in \cite{York:1972pr},\cite{York:1972sj}.
Reduction of degrees of freedom to lower dimensions also appeared later in the holography proposed in the black hole entropy context and more recently in AdS/CFT. 
A more mathematical search has also been conducted, e.g., in \cite{Isenberg,Fischer:1996qg,Gay-Balmaz:2014ena}.

 The Quantization carried out in \cite{Park:2014tia} was in the Lagrangian operator framework with inputs from the Dirac's Hamiltonian method.
In the present work, we discuss passing to the path integral description after reviewing and refining the analysis in \cite{Park:2014tia}.
We classically identify all the physical degrees of freedom, after which the canonical commutations relations can be imposed to quantize the system.  
The operator formalism, being on-shell, has certain advantages over the path integral approach.\footnote{By path integral, we mean the path integral that is referred to as the ``independent path-integral approach" below, i.e., the path integral approach not guided by the canonical operator analysis in precise identification of the physical degrees of freedom.}

Although it might be necessary to restrict the configuration space to globally hyperbolic spacetimes in order to have the reduction, such restriction must be only mild in nature, given that our goal at hand is perturbative analysis.
This is because small fluctuations around a given globally hyperbolic spacetime should all be globally hyperbolic since they must have the same topology.


The second half of this work is devoted to setting a stage for an explicit two-loop perturbative analysis in the three-dimensional description. Our 3D theory is not the genuine 3D gravity in that it inherits two physical degrees of freedom from the 4D theory: we are just using a ``3D window" to describe the 4D physics. Thus a legitimate perturbation theory can be set. The counterterms can be computed by using the background field method \cite{DeWitt:1975ys}\cite{Weinberg2}.

One-loop renormalizability was established in \cite{'tHooft:1974bx,Goroff:1985th} where the counterterms were obtained through the background field method. The metric was shifted according to $g_{\m\n}= \gh_{\m\n}+g^B_{\m\n}$ where  $g^B_{\m\n}$ is the background (or external) field. Once $\gh_{\m\n}$ was integrated out, a covariant expression resulted as it should. Although this procedure may be sufficient to establish the one-loop renormalizability, it is rather formal 
in that one would not precisely follow this procedure when one considers physics around a specific background. One would explicitly expand the action around a vacuum of interest (e.g., a flat spacetime), and consider perturbation around it. Although one faces the issue of background (in)dependence of renormalization in this approach, it should be a good starting point. As a matter of fact, the perturbative analysis around a flat background was carried out in \cite{Capper:1973pv} long ago.
We first review and extend the work of \cite{Capper:1973pv}. 
We also address issues that need to be understood to relate \cite{Capper:1973pv} to \cite{'tHooft:1974bx,Goroff:1985th}. This is the aforementioned issue of whether renormalizability depends on the choice of the background solution, and seems related to the question of what gauge invariant physical degrees of freedom are.

\vspace{.2in}
The rest of the paper is organized as follows. 
In the next section, we present a more detailed account of the holographic reduction observed in \cite{Park:2014tia}. We start with the Lagrangian and employ the Dirac's Hamiltonian method. The well-known non-dynamism of the lapse and shift leads then to the conclusion that only 3D residual symmetry is required for the removal of the shift vector. In other words, the residual symmetry after the bulk gauge-fixing (we use de Donder gauge) is sufficient to remove the shift vector. The removal of the shift vector renders the system three-dimensional. After review of the operator approach, we discuss the transition to the path integral. In section 3, we take the 3D description to prepare to carry out the two-loop renormalization.  First, we check the result of \cite{Capper:1973pv} obtained in the 4D context. Our results do not exactly match those of \cite{Capper:1973pv}. In our view, an error in \cite{Capper:1973pv} was made in applying a Ward-Takahashi type identity. We discuss what we believe to be the correct Ward-Takahashi identity. 
With our tools checked, we tackle the background (in)dependence of renormalizability in the current context. Next we turn to 3D.    
Unlike the genuine 3D gravity, the reduced system carries two physical degrees of freedom inherited from 4D; therefore it is legitimate to introduce the graviton propagator. We conclude with discussions and future directions.

\section{Constraints and quantization}
In this section, we review the quantization of 4D Einstein gravity proposed in \cite{Park:2014tia} with a more detailed and refined account. The operator method\footnote{See, e.g., the recent book by K. Huang \cite{Huang}.} - which is useful for carefully identifying the physical fields relevant for path integral quantization - will be considered first and a transition to the path integral will be discussed. The approach of \cite{Park:2014tia} contains elements from both the canonical and covariant methods - for example, it employs the ADM formalism on 
one hand but at the same time considers a fixed background and perturbative analysis around it. One of the salient features is solvability of the constraints that arise from the gauge-fixing of the lapse and shift in the Lagrangian.

The ADM formalism \cite{Arnowitt:1962hi} employs the 3+1 splitting
\bea
x^\m\equiv (y^m,x^3)
\eea
The separated-out coordinate $x^3$ will play the role of ``time" until the theory is reduced to 3D; afterwards the genuine time coordinate $t$ will be considered.\footnote{The unconventional splitting of ``time" and space appeared, e.g., in \cite{Sato:2002kv} in the supergravity context. It also appeared more recently in \cite{Hubeny:2011hd} in the fluid/gravity context; we thank J. de Boer for pointing out the use in this context.} 
The ``Hamiltonian of $x^3$ evolution" is obtained by applying the usual Legendre transformation to the ADM Lagrangian.
As well known, the Hamiltonian equations of motion reveal that the lapse and shift are non-dynamical. One can go from the classical theory to the quantum theory by imposing the canonical commutation relations. 
In particular, the equations of motion of the shift and lapse imply they are ``time"- (i.e., $x^3$-) independent.

The key step for the quantization is an elaborate gauge-fixing followed by solving of the resulting constraints: we will review that it is possible to gauge away the shift vector by exploiting the 3D symmetry left over after the bulk gauge-fixing by the 4D de Donder gauge.\footnote{This is one of the places where the analysis will be refined as compared to \cite{Park:2014tia}.} Although the relevance of a globally hyperbolic spacetime was discussed only in the more mathematical context of \cite{Park:2014qoa}, we will observe below how the potential relevance of a globally hyperbolic spacetime may arise in the context of \cite{Park:2014tia} as well.

\subsection{Dirac's method and Lagrangian quantization}

Consider the 4D Einstein-Hilbert action 
\bea
S=\int d^4 x \sqrt{-g}\;R  \la{unsplit}
\eea
Let us separate the $x^3$ out from the rest: 
\bea
x^\m\equiv (y^m,x^3)
\eea
where $\m=0,..,3$ and $m=0,1,2$. (See appendix A for the conventions.)
By parameterizing the 4D metric according to 
\bea
g_{\m\n}=\left(
\begin{array}{cc}
\g_{mn} & N_{ m} \\
&\\
N_{ n} &  n^2+\g^{mn}N_{m} N_{ n} 
\end{array}
\right)\quad,\quad
g^{\m\n}=\left(
\begin{array}{cc}
\g^{mn}+\fr1{n^2}N^m N^n & -\fr1{n^2}N^{ m} \\
&\\
-\fr1{n^2}N^{ n} &  \fr1{n^2} 
\end{array}
\right)
\eea
the 3+1 splitting yields (the boundary terms will not be kept track of) \cite{Arnowitt:1962hi}\cite{Misner,Poisson,Embacher:1995xp,Sato:2002kv}
\bea
S=\int d^4 x\;n\sqrt{-\g} \left(R^{(3)}+K^2-K_{mn}K^{mn}\right)
\la{1p3act}
\eea
with
\be
K_{mn}=\fr1{2n}\left(\mathscr{L}_{\pa_{x^3}} \g_{mn}-{\nabla}_m N_{n}
         -{\nabla}_n N_{ m} \right),\qquad K=\g^{mn}K_{mn}.
\la{K4defqq}
\ee
where $\mathscr{L}_{\pa_{x^3}}$ denotes the Lie derivative along the vector field $\pa_{x^3}$ and $\N_m$ is the 3D covariant derivative constructed out of $\g_{mn}$; $n$ and $N_{ m}$ denote the lapse function and shift vector respectively. The ``time" derivative does not act on $N_m$ or $n$ in their field equations, which read
\bea
{\N}_m (K^{mn}-\g^{mn} K)=0  \la{Ncon}
\eea
\bea
R^{(3)}-K^2+K_{mn}K^{mn}=0  \la{ncon}
\eea
We should take these as the constraints at the operator level. We come back to this Lagrangian system after going through the Hamiltonian formulation.
One can go to the Hamiltonian formulation by the usual Legendre transformation. The bulk part of the ``Hamiltonian" of $x^3$-evolution is
\bea
H=\int d^3y\left[ nh^{-1/2}({ -}\pi^{mn}\pi_{mn} { +} \fr12 \pi^2)
-n h^{1/2}R^\3-2N_m h^{1/2}\N_n(h^{-1/2}\pi^{mn})  \right]
\label{x3H}
\eea
where the canonical momentum is given by
\bea
\pi^{mn}=\sqrt{h}(K^{mn}-K h^{mn})
\eea
For any field $u$ including $n,N_m,\g_{mn}$, the Hamiltonian equation of motion can be written as
\bea
 \mathscr{L}_{\pa_{x^3}}u=[u,H]_P
\eea
where $[\cdots]_P$ denotes the Poisson bracket.
For $u=n,N_m$, the right-hand side vanishes without using any other relations such as the other constraints: 
\bea
\mathscr{L}_{\pa_{x^3}}{n}=0\quad,\quad \mathscr{L}_{\pa_{x^3}}{N}_m=0  \la{timeindep}
\eea
These equations tell that $n,N_m$ are non-dynamical. 
The fact that $n,N_m$ are $x^3$-independent can be explicitly used.

We now show that the dynamics associated with the $x^3$ evolution of the hypersurface is not genuine in the perturbative analysis around a vacuum compatible with the gauge-fixing that will be discussed shortly. As we will see, the shift vector can be gauge-fixed away by using the 3D residual symmetry. Nevertheless, there should be non-trivial dynamics within the hypersurface itself. Let us turn back to the Lagrangian formulation. 
Given the $x^3$-independence of $n,N_m$, the action takes
\bea
S=\int d^4 x\;n(y)\sqrt{-\g} \left(R^{(3)}+K^2-K_{mn}K^{mn}\right);
\la{1p3act2}
\eea
the constraints \rf{Ncon} and \rf{ncon} take the same forms
\bea
{\N}_m (K^{mn}-\g^{mn} K)=0  \la{Nconred}
\eea
\bea
R^{(3)}-K^2+K_{mn}K^{mn}=0  \la{nconred}
\eea
but now with
\be
K_{mn}=\fr1{2n(y)}\left(\mathscr{L}_{\pa_{x^3}} \g_{mn}-{\nabla}_m N_{n}(y)
         -{\nabla}_n N_{ m}(y) \right),\qquad K=\g^{mn}K_{mn}.
\la{K4defqq2}
\ee
In other words, we have explicitly used the fact that $n=n(y), N_m=N_m(y)$; $\g_{mn}$ still has the full $x^\m$-dependence. 
The bulk de Donder gauge $g^{\r\s}\G^\m_{\r\s}=0$ \cite{Smarr:1978dia} reads, in the ADM fields,
\bea
&& (\pa_{x^3}-N^m \pa_m) n=n^2K  \nn\\
&& (\pa_{x^3}-N^n \pa_n)N^m=n^2(\g^{mn}\pa_n \ln n-\g^{pq}\G^m_{pq})
\la{ADMdd}
\eea
This way of imposing the de Donder gauge differs in two respects from the usual way adopted in the perturbative analyses in the literature. Firstly, the linear form instead of the full form is used in perturbative analyses. Secondly, it is actually not the linear form itself but the square of the linear form so that the quadratic part of the action can be inverted to yield the propagator. We will impose the de Donder gauge as given in \rf{ADMdd}, and the usual form of de Donder gauge will be used in the next section where the perturbative analysis will be carried out. An analogous discussion with the axial gauge in a gauge theory can be found in \cite{Weinberg2}.
The gauge-fixing \rf{ADMdd} leaves residual symmetry that will play an important role. 
Let us consider the following to see the form of the residual symmetry. The non-covariant term in the 4D transformation of $g^{\r\s}\G^\m_{\r\s}$ is the second  term in the right-hand side of
\bea
g^{\r'\s'}\G^{\m'}_{\r'\s'}=\fr{\pa x^{\m'}}{\pa x^\m}\Big(g^{\r\s}\G^{\m}_{\r\s}\Big)-
g^{\r\s}\fr{\pa^2 x^{\m'}}{\pa x^\r\pa x^\s}
\eea
Let us set the non-covariant piece to zero:
\bea
g^{\r\s}\fr{\pa^2 x^{\m'}}{\pa x^\r\pa x^\s}=0 \la{resila}
\eea
For the case of an infinitesimal transformation $x^{\m'}=x^\m+\e^\m$, this becomes
\bea
g^{\r\s}\fr{\pa^2 \e^\m}{\pa x^\r\pa x^\s}=0 \la{resila2}
\eea
With this satisfied, the quantity $g^{\r\s}\G^\m_{\r\s}$ transforms as a contravariant vector. In other words, the transformation parameter that satisfies this condition will generate the residual symmetry. According to Theorem 10.1.2 of \cite{Wald}\footnote{The function $A$ in theorem 10.1.2 will need to be chosen as a metric dependent expression.} (see also Theorem 10.1.1), the equation \rf{resila2} admits a well-defined initial value problem (thus a solution) if the metric is that of a globally hyperbolic spacetime. As a matter of fact, all that is needed is for \rf{resila2} to admit a solution in order for the residual symmetry to exist: it might not be necessary to restrict the metric to that of a globally hyperbolic spacetime. However, at least as far as the perturbative analysis is concerned, a globally hyperbolic spacetime should not impose a serious restriction as stated before.
Since $N_m$ is three-dimensional, it should be possible, by choosing $\e^m$ (i.e., $\e^\m$ with $\e^3=0$) appropriately, to set 
\bea
N_m'=N_m+\N_3 \e_m^R=0  \label{ressym}
\eea
where the superscript $R$ stands for ``residual." This will set the initial conditions for $\e_m^R$.

Let us recapitulate: by using the residual 3D symmetry discussed above, one can gauge away the shift vector:
\bea
N_m=0
\eea
As shown in \cite{Park:2014tia} and \cite{Park:2014qoa}, the constraint equation \rf{Nconred} implies
\bea
\mathscr{L}_{\pa_{y^m}}n=0 \la{nconstr}
\eea
This, with the ``time"-independence of $n$, allows gauge-fixing $n=const$; for convenience we fix $n=1$, a valid choice for a flat background, the case of the main focus in this work.
With these fixings, the two equations of the de Donder gauge in \rf{ADMdd} simplify to
\bea
K=0  \la{Kz}
\eea
and
\bea
\g^{pq}\G^m_{pq}=0  \la{3ddd}
\eea
Note that \rf{3ddd} takes the form of the 3D de Donder gauge. This is only {\em apparently} true at this point: the arguments of $\g^{pq}$ and $\G^m_{pq}$ are still four-dimensional. However, this status will change.

Although the apparent 3D form of \rf{3ddd} may appear to be just a curiosity, it is more significant than this, as can bee seen from the following analysis. Let us take a $\g_{mn}$-variation of \rf{1p3act2} with the constraint \rf{nconred} taken into account:
\bea
\d S &=&\d\int d^4 x\;\sqrt{-\g} \left(R^{(3)}+K^2-K_{mn}K^{mn}\right)\nn\\
 &=&\int d^4 x\;(\d\sqrt{-\g}) \left(R^{(3)}+K^2-K_{mn}K^{mn}\right)
    + \left(\d R^{(3)}+\d[K^2-K_{mn}K^{mn}]\right)\nn\\
\eea
Upon substituting $K^2-K_{mn}K^{mn}=R^{(3)}$ and $\d[K^2-K_{mn}K^{mn}]=\d R^{(3)}$, one gets
\bea
\d S =2\d \int d^4 x\;\sqrt{-\g} R^{(3)}
\eea
Therefore, the action takes the three-dimensional form other than the measure and the implicit $x^3$-dependence of $R^{(3)}$. Let us omit the factor 2: 
\bea
 S =\int d^4 x\;\sqrt{-\g} R^{(3)}
\eea
Its field equation is 
\bea
R^\3_{mn}=0\quad,\quad R^\3=0  \la{3dmeom}
\eea
On account of \rf{3dmeom} and \rf{Kz}, the constraint \rf{nconred} becomes
\bea
K_{mn}K^{mn}=0
\eea
This implies, up to the Wick rotation in the time $t$ direction,
\bea
K_{mn}=0
\eea
which, in turn, implies reduction of the metric at the operator (and classical) level,
\bea
\g_{mn}(x^\m)\Rightarrow \g_{mn}(y^p)
\eea
The upshot of the operator analysis so far is that, for the perturbative analysis around a vacuum on which \rf{nconstr} can be imposed (Minkowski spacetime is a representative example), the original system reduces to
\bea
 S = \int d^4 x\;\sqrt{-\g} R^{(3)} \la{eff3Dact}
\eea
with the reduced gauge-fixing condition
\bea
\g^{pq}\G^m_{pq}=0  \la{3dddz}
\eea
As shown in \cite{Park:2014tia} and presented for review in the next section, the action effectively further reduces to
\bea
 S = \int d^3 x\;\sqrt{-\g} R^{(3)}
\eea
with renormalization of the Newton's constant that has been suppressed. 
The analysis so far implies that the path integral corresponding to the Lagrangian operator analysis takes
\bea
<\mbox{Vac},out|\mbox{Vac},in>=
\int d\g_{rs} \;  e^{i\int d^4x R^\3}             \det({\cal F})\; \d(f)\d\Big[\g^{pq}\G^m_{pq}\Big]
\la{pigc}
\eea
where the $\det({\cal F})$ is the Faddeev-Popov determinant associated with the gauge-fixing $\d\Big[\g^{pq}\G^m_{pq}\Big]$. ($f$ will be chosen as the traceless condition of the 3D metric below; this choice does not introduce a non-trivial determinant.)
This completes the discussion of the Lagrangian operator approach and the corresponding path integral description. In the next subsection, we side-step to examine a more path integral-centered approach and reach the same conclusion.

\subsection{``independent" path integral approach}
In the previous subsection, we started with the canonical operator analysis, and, thus guided, could write the path integral down in \rf{pigc}. We will use \rf{pigc} in the next section to carry out an explicit renormalization procedure. In the present subsection, we sidestep to explore a more path integral-centered attempt to quantize gravity.

If is often stated in the general gauge theory context in literature that one may take the path integral, instead of the canonical analysis, as the starting point (hence the title of this subsection). We adopt this viewpoint in this subsection and push quantization as far as possible. As we will see, this exercise leads to a result not fully consistent with the path integral approach guided by the canonical operator analysis, namely \rf{pigc}. We attribute this discrepancy to the fact that the unphysical shift vector is not gauge-fixed in this ``independent" path integral approach.

Let us first examine how the gauge symmetry is realized in the ADM formulation.
Early work on gauge symmetry in the ADM Hamiltonian formalism can be found in \cite{Bergmann:1972ud}. The symmetry generated by the shift vector constraint, the so-called spatial diffeomorphism, takes a simple form; it is the symmetry generated by the lapse function constraint that has complications. 
The present analysis should shed light on the origin of the complication of the conventional Hamiltonian analysis, and will be used in the subsequent discussion of the path integral description.

The action of the ADM formulation has manifest 3D gauge invariance with the 4D invariance not as manifest. In terms of the ADM variable the 4D transformation is a non-linear field-dependent transformation since the lapse function is given by a non-linear field redefinition of the usual metric components. 
The usual infinitesimal coordinate transformation, $\d g_{\m\n}=\N_\m\e_\n+\N_\n\e_\m$, can be translated into the rules in the ADM formalism. The shift vector transformation is the same as $g_{3m}$; the lapse function transformation can be determined from
\bea
\d g_{33}=\d(n^2+\g^{mn}N_{m} N_{ n})=2n\d n+2 N_m\g^{mn}\d N_n
       +N_m (\d \g^{mn}) N_n
\eea
Solving this for $\d n$, one gets
\bea
\d n &=& \fr{\d g_{33}-2 N_m\g^{mn}\d N_n-N_m (\d \g^{mn}) N_n}{2n}
\eea
The complexity seen in \cite{Bergmann:1972ud} should be related to the fact that $\d n$ takes a rather complicated form.\footnote{
Although $\d n$ has a highly non-linear and complicated metric dependence,
it becomes simple once the gauge conditions $n=1, N_m=0$ are imposed:
\bea
\d n|_{n=1,N_p=0} =\fr12 \d g_{33}= {\N_3 \e_3}|_{n=1,N_p=0}=\pa_3 \e_3
\eea
The transformation parameter $\e_\m$ with the $\e_3=0$ will therefore preserve $n$, and this is consistent with our finding above \rf{ressym}.
}
Let us turn to the ``independent" path integral.
It is rather obvious from the beginning
that staying entirely within the path integral formulation (i.e., without the close guidance from the canonical operator formalism) should be impossible since, for one thing, the physical states must be determined through the operator description.
One may still try to construct as completely as possible the path integral expression that would be the counterpart of \rf{pigc}, and see where this approach leads. As we will show now, this approach points towards the same direction as the operator approach to a certain extent. However, the resulting expression does not completely coincide with \rf{pigc}, and the reason will be found in gauge-{\em unfixing} of the shift vector.

One's first guess for the path integral expression would be\footnote{We will comment shortly on the relationship between this path integral and the more covariant form $\int dg_{\m\n}(\cdots)e^{i\int d^4 x R}$ where $(\cdots)$ denotes the gauge-fixing related terms. \la{cm}}
\bea
&&<\mbox{Vac},out|\mbox{Vac},in>
=\int dn dN_{l_1} d\g_{l_2l_3} d\pi^{l_4l_5} \;  e^{i\int (\dot{\g}_{rs} \pi^{rs} -{\cal H})}             \det({\cal D})\nn\\
&& \d\Big[(\pa_{y^3}-N^p \pa_p) n-n^2K  \Big]  
\d\Big[(\pa_{y^3}-N^q \pa_q)N^m-n^2(\g^{mq}\pa_q \ln n-\g^{pq}\G^m_{pq})\Big]\nn\\
\label{1stguess}
\eea
where  ${\cal H}$ is the Hamiltonian density, $H=\int d^3 y {\cal H}$; the arguments of the delta functions are the de Donder gauge conditions (i.e., $\g^{\r\s}\G^\m_{\r\s}=0$) in the ADM variables, and $\det({\cal D})$ denotes the determinant factor that corresponds to this gauge.

However, there are several undesirable features in the expression above.
For example, there is an ambiguity with regards to whether one should use $K_{mn}$ (or equivalently $\pa_{x^3} \g_{mn}$) or $\pi_{mn}$ inside the delta functions. To get the usual Lagrangian after the momentum integration, one should use $K_{mn}$ (as indicated in \rf{1stguess}) and treat the momenta $\pi_{mn}$ independently of the $K$-fields when performing the $\pi^{mn}$ integration. More seriously, the path integral form above leads to a clash between the lapse/shift constraint and the de Donder gauge conditions:
The constraints ${\N}_m (K^{mn}-\g^{mn} K)=0,R^{(3)}-K^2+K_{mn}K^{mn}$ have not been imposed; we would like them to be automatically imposed by the path integral over $n,N_m$. But then, one encounters a problem since $n$ and $N_m$ have appeared inside the delta function.\footnote{There is another fact that should be taken into account. The coefficient of $\pi_{mn}^2$ in \rf{x3H} is field dependent; therefore, the $\pi_{mn}$ integration will produce a determinant factor. Such a factor can be removed by introducing an appropriate determinant factor in the path integral measure. This can also be viewed as part of the renormalization procedure \cite{Honerkamp:1996va,Honerkamp:1971sh,Tseytlin:1989si}.}

Because of these factors, one is led to consider a gauge condition that does not involve $K_{mn}$ or $n,N_m$. As far as we can see, this measure (i.e., employing a gauge condition not involving $K_{mn}$ or $n,N_m$) should be the only reasonable thing to correctly enforce the lapse and shift constraints. Motivated in part by the operator analysis as well, let us consider
\bea
<\mbox{Vac},out|\mbox{Vac},in>
=\int dn dN_{l_1} d\g_{l_2l_3} d\pi^{l_4l_5} \;  e^{i\int (\dot{\g}_{mn} \pi^{mn} -{\cal H})}             \det({\cal F})\;  
\d(f)\d\Big[\g^{pq}\G^m_{pq}\Big]\nn
\eea
where the $\det({\cal F})$ is the Faddeev-Popov determinant factor associated with the ``3D" de Donder gauge $\d\Big[\g^{pq}\G^m_{pq}\Big]$. ($f$ will be chosen as before in the previous subsection.)
The path integral over the momenta will force the field equation that relates the canonical fields and their conjugate momenta; thus we may freely insert the delta function with this field equation as the argument: 
\bea
&&<\mbox{Vac},out|\mbox{Vac},in>\nn\\
\!\!\!\!\!\!=&&\!\!\!\!\!\!\int dn dN_l d\g_{rs} d\pi^{mn} \;  e^{i\int (\dot{\g}_{mn} \pi^{mn} -{\cal H})}             \det({\cal F})\d\Big[ \pi^{mn}-\sqrt{h}(K^{mn}-K h^{mn})\Big] \d(f)\d\Big[\g^{pq}\G^m_{pq}\Big]\nn
\eea
Let us carry out the $n,N_m$ integration; the path integral takes
\bea
&&<\mbox{Vac},out|\mbox{Vac},in>\nn\\
=&&\int d\g_{rs} d\pi^{mn} \;  e^{i\int (\dot{\g}_{mn} \pi^{mn} )}             \det({\cal F})\d\Big[ \pi^{mn}-\sqrt{h}(K^{mn}-K h^{mn})\Big]\;
 \d(f)\d\Big[\g^{pq}\G^m_{pq}\Big]\nn\\
=&&\int d\g_{rs} \;  e^{i\int (\dot{\g}_{mn} [\sqrt{h}(K^{mn}-K h^{mn})] )}             \det({\cal F})\; \d(f)\d\Big[\g^{pq}\G^m_{pq}\Big]
\la{cpi}
\eea
The lapse and shift constraints are to be imposed separately:
\bea
{\N}_m (K^{mn}-\g^{mn} K)=0  
\eea
\bea
R^{(3)}-K^2+K_{mn}K^{mn}=0  
\eea
{A direct integration over $n,N_m$ would put these in the path integral in the form of delta functions. Separate consideration of them, i.e., outside of the path integral, is reminiscent of old covariant quantization of string theory.}
With the steps above, the $e^{i\int d^4x (\dot{\g}_{mn} \pi^{mn} )}$ factor becomes $e^{i\int d^4x R^\3}$, bringing the path integral close to the operator formulation:
\bea
<\mbox{Vac},out|\mbox{Vac},in>=
\int d\g_{rs} \;  e^{i\int d^4x R^\3}             \det({\cal F})\; \d(f)\d\Big[\g^{pq}\G^m_{pq}\Big]
\eea
Let us turn to the issue stated in footnote \ref{cm}. As often stated in the gauge theory context, one may take, as the starting point, the more covariant form of the path integral in terms of the usual metric component measure
\bea
\int dg_{\m\n}\;(\cdots)e^{i\int d^4 x\;R} \la{mcpi}
\eea
where $(\cdots)$ represents the gauge-fixing conditions and corresponding Faddeev-Popov determinants: the de Donder gauge and $n=1,N_m=0$ with the corresponding constraints (with everything expressed in terms of $g_{\m\n}$).
The Jacobian for the change of variables from $g_{\m\n}$ to $(n,N_m, \g_{mn})$ is $2n$:
\bea
d g_{\m\n}=dn dN_m d\g_{pq}\,(2n)
\eea
The determinant factor from the change of the variables becomes 1 (2 more precisely) for the $n=1$ gauge-fixing. Also, the Faddeev-Popov determinant factor becomes field-independent, thus immaterial. This way, it seems possible to make a connection between the current approach and the more covariant approach. 
In particular, the measure in \rf{pigc} can be viewed as a gauge-fixed version of the 4D diffeomorphism invariant measure.

\begin{figure}[tbp]
\centering 
\includegraphics[width=.8\textwidth,trim=0 530 0 80,clip]{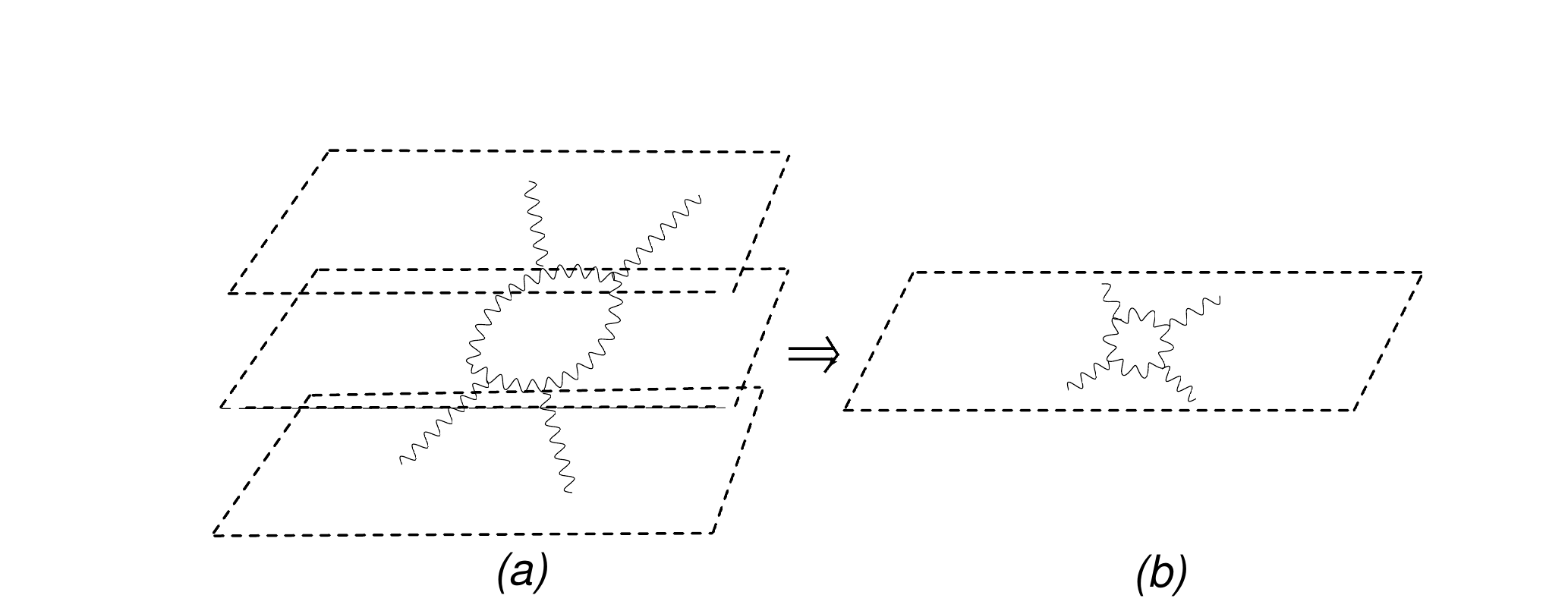}
\caption{ (a) 4D scattering   (b) projection onto 3D hypersurface}
\end{figure}

\section{Effectively 3D perturbative analysis}
By the canonical operator quantization in the previous subsections, we have precisely identified the physical fields and deduced the path integral expression.
A legitimate perturbative analysis can be set up with \rf{pigc}, and for the gauge-fixing, the usual linearized version of \rf{3dddz} will be imposed. 
The renormalization procedure involves much tedious algebra that is best handled by computer codes. We employ a Mathematica package xAct`xTensor`. Since the number of terms is quite large, we have decide to check our strategy and apparatus against the known results first; namely, the 4D analysis in \cite{Capper:1973pv} in which one-loop divergence of two-point amplitude was computed. 
As we will show below our result does not fully agree with \cite{Capper:1973pv}. We believe that the reason for the discrepancy is their use of an erroneous Ward-Takahashi type identity, while our result satisfies the correct Ward-Takahashi type identity.

The counterterms were not computed in \cite{Capper:1973pv}; we go further and compute the explicit counterterms by applying the background field method to the action expanded around a flat background.
This procedure is different from that of \cite{'tHooft:1974bx} in that we {\em do} expand the action around a specific background (i.e., a flat background in the present case). Although the type of analysis carried out in \cite{'tHooft:1974bx} may be sufficient to establish renormalizability,\footnote{We will come back to this point later.} one will have to fix a background in order to study physical processes such as scattering of gravitons in a given background.

The difference in the approaches of \cite{'tHooft:1974bx} and the present paper will pose a subtle question; let us first contrast the two approaches. 
We illustrate the issue by taking the scalar $\l \z^3$ theory,
\bea
\cL=-\fr12\pa_\m \z \pa^\m \z-\fr{\l}{3!}\z^3  \la{ztheory}
\eea
Consider the trivial vacuum $\z=0$ and the perturbation theory around it. 
One can compute the counterterms by the background field method in which the field is
shifted 
\bea
\z\ra \z+\z_B
\eea
Once one integrates out $\z$ in such a way that $\z_B$ becomes external lines, one can obtain the counterterms for the one-loop two-point diagram.  
One may also consider a nonzero constant vacuum $\z=\z_0$ and expand the theory \rf{ztheory} around this vacuum. 
\bea
\cL=-\fr12\pa_\m \z \pa^\m \z-\fr{\l}{3!}(\z+\z_0)^3  \la{z0theory}
\eea
This theory will have one-loop divergence, and one can compute the counterterms again by employing the background field method; shift the action \rf{z0theory} around $\z=\z_B$
\bea
\cL=-\fr12\pa_\m (\z+\z_B) \pa^\m (\z+\z_B)-\fr{\l}{3!}(\z+\z_B+\z_0)^3  \la{z0theoryde}
\eea
and integrate out the $\z$ field in such a way that the background field becomes external lines. Both with \rf{ztheory} and \rf{z0theory}, it will be possible to absorb the counterterms into the existing terms in the shifted actions, and the background independence of renormalization is demonstrated.\footnote{The author thanks M. Rocek for the discussions on related issues.} We also have explicitly checked this for the diagram Fig. 3 (a) below in which the lines now represent the gauge fields.

\begin{figure}[tbp]
\centering 
\includegraphics[width=1\textwidth,trim=50 490 0 75,clip]{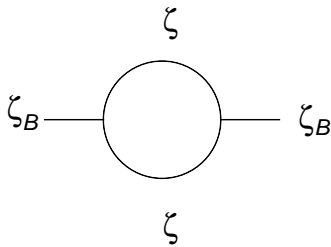}
\caption{background method for scalar theory}
\end{figure}

The background actions in \cite{'tHooft:1974bx,Deser:1974cy,Deser:1974cz,Goroff:1985th} were obtained 
by shifting the metric $g_{\m\n}= \gh_{\m\n}+g^B_{\m\n}$ where  $g^B_{\m\n}$ is the background (or external) field and integrating out $\gh_{\m\n}$. The resulting background action is covariant.\footnote{The definition of 1PI action through the usual background method gives the effective action when the field is expanded around zero vev. In other words, if it is possible to do perturbation theory around $g_{\m\n}=0$, the counterterms obtained in \cite{'tHooft:1974bx} would be appropriate.} The perturbation of the scalar theory around $\z=0$ above is analogous to this. However, there is a big difference between non-gravity theories and gravity theories. For a gravity theory, perturbation around $g_{\m\n}=0$ is not defined.

In this work, we consider shifting the metric $\gt^{\m\n}= \f^{\m\n}+\gt_0^{\m\n}$ \footnote{The use of $\gt^{\m\n}$ instead of $g^{\m\n}$ is irrelevant for the issue at hand.} where $\gt^{\m\n}\equiv \sqrt{-g}g^{\m\n}$ and $\gt^0_{\m\n}$ (or more precisely, $g^0_{\m\n}$) is a solution of the Einstein equation (such as a flat metric) to obtain the expanded action as in \cite{Capper:1973pv}. Then we shift $\f^{\m\n}$ itself according to $\f^{\m\n}\ra  g_B^{\m\n}+\f^{\m\n}$ where  $g^B_{\m\n}$ is the background (or external) field with the resulting action and obtain the doubly expanded action. We integrate $\f^{\m\n}$ out and obtain the background effective action. The situation is analogous to the perturbation around $\z=\z_0$ above, but unlike the scalar theory, the counterterms cannot be reabsorbed into the existing terms in the action or field redefinition of the metric. The resulting action is not covariant although it is the correct counter term action. In other words, we employ the background field method according to ch.16 of \cite{Weinberg2} in order to conveniently produce the counterterms for the action expanded around a flat background. Since the action is expanded around a specific background first, the counterterms are not expected to have covariant forms, and they indeed turn out to be non-covariant. 
Since the one-loop divergence and counterterms do not vanish, this brings us back to the question of one-loop renormalizability of the 4D action, an issue  that we take up with other related ones in Discussions. We put these subtle issues aside for the moment and carry out the computations of divergences and their counterterms in the next subsection.

\subsection{review of 4D case}
In terms of $\gt^{\a\b}\equiv \sqrt{-g}g^{\a\b}$, the Einstein-Hilbert action reads
\bea
S&=&-\int d^Dx \fr14\Big(
\gt_{\k_1\k_3}\gt_{\k_2\k_4}  \gt^{\a\b}\pa_\a \gt^{\k_1\k_2}\pa_\b \gt^{\k_3\k_4} 
-2 \gt_{\k_1\k_2}\pa_\a \gt^{\k_1 \b} \pa_\b \gt^{\k_2 \a}\nn\\ 
&&\hspace{1.5in}-\fr1{D-2}\gt^{\a\b}\gt_{\k_1\k_2}\pa_\a \gt^{\k_1\k_2}
\gt_{\k_3\k_4}\pa_\b \gt^{\k_3\k_4}
\Big)
\eea
where $D$ denotes the spacetime dimensions and will be set $D=4$ in this subsection. The advantage of using ${\gt}^{\m\n}$ over $g^{\m\n}$ is that the number of cubic couplings is substantially reduced.
Let us numerically rescale this action by a factor 2 so that
with
\bea
\gt^{\m\n}\equiv \eta^{\m\n}+\f^{\m\n}
\eea
the quadratic and cubic order actions are given by
\bea
{\cL}_\2 &\equiv& 
\int -\fr12\pa_{\a}\f_{\m\n}\pa^{\a}\f^{\m\n}
+ \f^\k \f_\k+\fr1{2(D-2)}\pa_\m \f\pa^\m \f \nn\\
{\cL}_\3 &\equiv&
\int -\fr12\f^{\a\b} \pa_\a \f^{\r\s}\pa_\b \f_{\r\s}
+\f_{\r\s}\pa^\a \f^{\r\k}\pa_\a \f_{\k}^\s
-\f_{\r\s}\pa_\a \f^{\r\b}\pa_\b \f^{\s\a} \nn\\
&&\qquad -\fr1{2(D-2)}\Big(-\f^{\a\b}\pa_\a\f \pa_\b \f 
+2\f_{\r\s}\pa_\a \f \pa^\a \f^{\r\s}\Big)  \label{cubicterms}
\eea
where we have introduced short-hand notations
\bea
\f\equiv \eta_{\m\n}\f^{\m\n}\quad,\quad \f^\m\equiv \pa_\k\f^{\k\m}
\eea
The gauge-fixing term is
\bea
\cL_{g.c.}=-\f_\m\f^\m
\eea
and with this the quadratic piece becomes
\bea
\cL_\2+\cL_{g.c.}= -\fr12\pa_{\a}\f_{\m\n}\pa^{\a}\f^{\m\n}
+\fr1{2(D-2)}\pa_\m \f\pa^\m \f 
\eea
The 4D propagator is given by
\bea 
<\f_{\m\n}(x_1)\f_{\r\s}(x_2)>= P_{\m\n\r\s} \int \fr{d^4k}{(2\pi)^4}\fr{e^{ik\cdot (x_1-x_2)}}{i k^2}
\eea
where
\bea
P_{\m\n\r\s}\equiv \fr12(\eta_{\m\r}\eta_{\n\s}+\eta_{\m\s}\eta_{\n\r}- \eta_{\m\n}\eta_{\r\s})
\eea

\begin{figure}[tbp]
\centering 
\includegraphics[width=.9\textwidth,trim=0 620 0 65,clip]{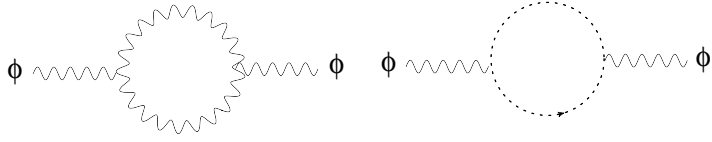}
\caption{(a) graviton loop  (b) ghost loop}
\end{figure}

One can show (from (4.6) of \cite{Fradkin:1970pn}) that the quadratic and cubic ghost terms are given by
\bea
-\pa_\m \Cb_\n\, C_\b\pa^\b \gt^{\m\n}
+\pa_\n \Cb_\m \pa^\n C_\g \gt^{\m\g}
+\pa_\n \Cb_\m \pa_\g C^\m \gt^{\n\g}
-\pa_\m \Cb_\n\pa^\b  C_\b\gt^{\m\n}
\eea
which, upon substituting $\gt^{\m\n}\equiv \eta^{\m\n}+\f^{\m\n}$, yields
\bea
\cL_{\2}^{gh}+\cL_{\3}^{gh}&\equiv&\Cb^\r \pa^\s \pa_\s C_\r\;\;-\Cb_\r (\pa_\k\pa^\s \f^{\r\k}) C_\s
+\Cb^\r (\pa_{\k_1} \f^{\k_1\k_2}) \pa_{\k_2}C_\r
\nn\\
&&
\hspace{.2in}-\Cb_{\r} (\pa_{\k_1} \f^{\r\k_1}) \pa_{\k_2}C^{\k_2}+\Cb^\r \f^{\k_1\k_2}\pa_{\k_1} \pa_{\k_2}C_\r  
\eea
Let us partially integrate and obtain the following form more convenient for Mathematica manipulations:
\bea
\cL_{\3}^{gh}&\equiv&-\f^{\r\k_2} \pa_{\k_1} \Cb_\r \pa_{\k_2}C^{\k_1}
-\f^{\k_1\k_2} \pa_{\k_1} \Cb_\r \pa_{\k_2}C^{\r}
+\f^{\r\s} \pa_{\r} \Cb_\s \pa_{\k}C^{\k}\nn\\
&&\hspace{1.5in} 
+\pa_\k\f^{\r\s} \pa_{\r} \Cb_\s\, C^{\k}
\label{ghterms}
\eea
The ghost propagator is given by
\bea
<C_\m(x_1)\Cb_\n(x_2)>=\eta_{\m\n}\int
 \fr{d^4k}{(2\pi)^4}\;\fr{1}{ik^2}e^{ik\cdot(x_1-x_2)}
\eea
\ssk \ssk
\ni For the 1-loop diagram Fig. 3, one should consider
\bea
<\f_{\m\n}(x_1)\f_{\r\s}(x_2)\Big(-\fr12\Big)\Big(\int {\cL}_\3\Big)^2 >_{1-loop}, 
\la{grav1loop}
\eea
a tedious computation.
The computation can be best handled by a computer and we employ the Mathematica package xAct`xTensor`. The package can conveniently be used, once the cubic coupling 
is written
\bea
{\cL}_\3 &\equiv& \f_{\l_1\l_2}\pa_{\a_1}\f_{\l_3\l_4}\pa_{\b_1}\f_{\l_5\l_6}
T^{\a_1\b_1\l_1\l_2\l_3\l_4}
\eea
where
\bea
&& T^{\a_1\b_1\l_1\l_2\l_3\l_4}\equiv -\fr12 \eta^{\l_1\a_1}\eta^{\l_2\b_1}\eta^{\l_3\l_5}\eta^{\l_4\l_6}
+\eta^{\l_1\l_3}\eta^{\l_2\l_6}\eta^{\a_1\b_1}\eta^{\l_4\l_5}
\nn\\
&& -\eta^{\l_1\l_3}\eta^{\l_2\l_5}\eta^{\a_1\l_6}\eta^{\b_1\l_4}
+\fr14 \eta^{\l_1\a_1}\eta^{\l_2\b_1}\eta^{\l_3\l_4}\eta^{\l_5\l_6}
-\fr12\eta^{\l_1\l_5}\eta^{\a_1\b_1}\eta^{\l_3\l_4}\eta^{\l_2\l_6} \nn
\eea
Then \rf{grav1loop} can be rewritten
\bea
\!\!\!\!\!\!\!\!\!\!\!\!-\fr12\int \int <\f_{\m\n}(x_1)\f_{\r\s}(x_2)\f_{\l_1\l_2}\pa_{\a_1}\f_{\l_3\l_4}\pa_{\b_1}\f_{\l_5\l_6}
\f_{\l_1'\l_2'}\pa_{\a_1'}\f_{\l_3'\l_4'}\pa_{\b_1'}\f_{\l_5'\l_6'}>
T^{\a_1\b_1\l_1\l_2\l_3\l_4}T^{\a_1'\b_1'\l_1'\l_2'\l_3'\l_4'}\nn
\eea
After one computes 
\bea
<\f_{\m\n}(x_1)\f_{\r\s}(x_2)(\f_{\l_1\l_2}\pa_{\a_1}\f_{\l_3\l_4}\pa_{\b_1}\f_{\l_5\l_6})
(\f_{\l_1'\l_2'}\pa_{\a_1'}\f_{\l_3'\l_4'}\pa_{\b_1'}\f_{\l_5'\l_6'})>_{1-loop}
\eea
and goes to the momentum space, the multiplication by $T^{...}T^{...}$ can effectively be carried out by xAct`xTensor`.  
The following result is obtained for the graviton sector:
\bea
&&\int \fr{d^4x_1}{(2\pi)^4}e^{ip_1\cdot x_1}\int \fr{d^4x_2}{(2\pi)^4}e^{ip_2\cdot x_2}<\f_{\m\n}(x_1)\f_{\r\s}(x_2)>_{grav\; 1-loop}\nn\\
&{\Rightarrow}&(2\pi)^4\d(\Sigma p_i)\fr{\G(\e)}{(4\pi)^2}\Big[-\frac{17}{48} {p_1}^4 (\eta^{ \m \s} \eta^{\n \r}+ \eta^{\m \r} \eta^{\n \s})-\frac{7}{16} {p_1}^4
   \eta^{\m\n} \eta^{\r \s}\nn\\
 &&+\frac{1}{3} {p_1}^2(
   {p_1}^\n {p_1}^\s \eta^{\m \r}+ {p_1}^\n {p_1}^\r \eta^{\m \s}+ {p_1}^\m {p_1}^q \eta^{\n \r}+ {p_1}^\m {p_1}^\r \eta^{\n \s})\nn\\
 &&+\frac{19}{24} {p_1}^2 ({p_1}^\r {p_1}^\s \eta^{\m\n}+ {p_1}^\m {p_1}^\n \eta^{\r \s})-\frac{11}{6} {p_1}^\m {p_1}^\n{p_1}^\r {p_1}^\s \Big]
\eea
where $\e\equiv 2-D/2$ and $\Rightarrow$ indicates
\bea
{\Rightarrow}: \mbox{divergent part of the 1PI diagram}
\eea
Not all of the coefficients match those obtained in \cite{Capper:1973pv}.
(The coefficient of ${p_1}^2 {p_1}^p {p_1}^q \eta^{mn}$, $\frac{19}{24}$, for example, does not match.) We will come back to this point after discussing the ghost sector in which a similar discrepancy is found. 
Our result for the ghost sector is
\bea
&&\int \fr{d^4x_1}{(2\pi)^4}e^{ip_1\cdot x_1}\int \fr{d^4x_2}{(2\pi)^4}e^{ip_2\cdot x_2}<\f_{\m\n}(x_1)\f_{\r\s}(x_2)>_{ghost\;1-loop} \nn\\
&\Rightarrow& (2\pi)^4\d(\Sigma p_i)\fr{\G(\e)}{(4\pi)^2}\Big[\frac{1}{60} {p_1}^4 (\eta^{\m \s} \eta^{\n \r}+\eta^{\m \r} \eta^{ \n \s})+\frac{59}{240} {p_1}^4
   \eta^{\m\n} \eta^{ \r\s}\nn\\
&&+\frac{1}{240} {p_1}^2
   ({p_1}^\n {p_1}^\s \eta^{ \m \r} + {p_1}^\n {p_1}^\r \eta^{\m \s}+ {p_1}^\m {p_1}^\s \eta^{ \n \r}+ {p_1}^\m {p_1}^\r \eta^{\n \s})\nn\\
 &&-\frac{7}{20} {p_1}^2 ({p_1}^\r {p_1}^\s \eta^{ \m\n} + {p_1}^\m {p_1}^\n \eta^{\r \s})+\frac{7}{15} {p_1}^\m {p_1}^\n
   {p_1}^\r {p_1}^\s  \Big]
\eea
This result again is different from that of \cite{Capper:1973pv}. The coefficient of $ {p_1}^4\eta^{\m\n} \eta^{ \r\s}$, $\frac{59}{240}$, for example, is different from the corresponding result in \cite{Capper:1973pv}.
We manually double-checked the coefficient of $ {p_1}^4\eta^{\m\n} \eta^{\r\s}$ by computing $<\f^{11}\f^{22}>$ with the condition $p_1^{\m=1}=0=p_1^{\m=2}$, which substantially reduces the amount of algebra; the manual computation confirmed the coefficient $\frac{59}{240}$.
The sum of the graviton and ghost contributions is given by
\bea
&&\int \fr{d^4x_1}{(2\pi)^4}e^{ip_1\cdot x_1}\int \fr{d^4x_2}{(2\pi)^4}e^{ip_2\cdot x_2}<\f_{\m\n}(x_1)\f_{\r\s}(x_2)>_{1-loop\;total}\nn\\
&\dot{=}&(2\pi)^4\d(\Sigma p_i)\fr{\G(\e)}{(4\pi)^2}\Big[-\frac{27}{80} {p_1}^4 (\eta^{\m\s} \eta^{\n\r}+ \eta^{\m\r} \eta^{\n\s})-\frac{23}{120}
   {p_1}^4 \eta^{\m\n} \eta^{\r\s}\nn\\
&&   +\frac{27}{80}
   {p_1}^2( {p_1}^\n {p_1}^\s \eta^{\m\r}+ {p_1}^\n {p_1}^\r \eta^{\m\s}+ {p_1}^\m {p_1}^\s \eta^{\n\r}+ {p_1}^\m
   {p_1}^\r \eta^{\n\s})\nn\\
&& +\frac{53}{120} {p_1}^2 ({p_1}^\r {p_1}^\s \eta^{\m\n}  + {p_1}^\m {p_1}^\n \eta^{\r\s})-\frac{41}{30} {p_1}^\m
   {p_1}^\n {p_1}^\r {p_1}^\s  \Big]
   \la{totaldiv}
\eea
The discrepancy observed above seems to be due to the application of an incorrect identity
in \cite{Capper:1973pv} wherein it was stated that the following expression is independent of $B^\n$:  
\bea
Z[j_{\m\n}]=\int d\gt^{\r\s}\D[\gt^{\a\b}]\d(\pa_\m\gt^{\m\n}-B^\n)
e^{i\int \cL+j_{\k_1\k_2}\gt^{\k_1\k_2}}
\eea
based on which a Ward-Takahashi type identity was written.
The expression actually {\em does} depend on the gauge-fixing term $\d(\pa_\m\gt^{\m\n}-B^\n)$ (and thus on $B^\n$) through a field-independent factor \cite{Weinberg2}. For the correct identity, let us consider
\bea
Z[j_{\m}]=\int d\f^{\r\s}\det({\cal F})e^{-i\int\f^{\k}\f_\k}e^{i\int \cL_\2+\cL_\3+\cdots+ j_{\n}\,\f^{\n}}
\eea
This is the path integral that we have been using other than the presence of 
$\f^\n j_\n$. This path integral depends on $j_\n$; schematically the dependence goes as $\sim e^{j(\cdots)j}$. We have checked that the graviton 3-point amplitude $<\f_\m\f_\n\f_\r>$ at tree-level vanishes.\footnote{ There seems to be some kind of a non-renormalization theorem:
 The following two-point function at one-loop vanishes:
\bea
<\f^\m\,\f^\n>=0
\eea
(We checked that our result \rf{totaldiv} satisfies the corresponding momentum space version of this: once the expression in \rf{totaldiv} is contracted with $p_1^\n p_2^\s$, the resulting expression vanishes.)}
(The correlator $<\f_\m\f_\n\f_\r>$ should not vanish if the Ward-Takahashi type identity of \cite{Capper:1973pv} is correct.)

\ssk\ssk\ssk

Above, we have computed the total one-loop divergence in the two-point amplitude.
Now we turn to the corresponding counterterms.
We illustrate the computations involved with a sample calculation in the ghost sector; the computation for the rest of the ghost sector and graviton sector can be found in one of the appendices. 
The cubic terms that involve the ghost fields are
\bea
\cL_{\f CC}\equiv \cL_{\3}^{gh}\equiv -\f^{\r\k_2} \pa_{\k_1} \Cb_\r \pa_{\k_2}C^{\k_1}
-\f^{\k_1\k_2} \pa_{\k_1} \Cb_\r \pa_{\k_2}C^{\r}
+\f^{\r\s} \pa_{\r} \Cb_\s \pa_{\k}C^{\k}
+\pa_\k\f^{\r\s} \pa_{\r} \Cb_\s\, C^{\k}
\label{ghtermscubic}
\nn\\
\eea
Let us shift the field 
\bea
\f^{\m\n}\ra \f^{\m\n}+\vf^{\m\n}
\eea
where $\vf^{\m\n}$ represents the background field.
The relevant part of the shifted action is
\bea
\cL_{\vf CC}\equiv -\vf^{\r\k_2} \pa_{\k_1} \Cb_\r \pa_{\k_2}C^{\k_1}
-\vf^{\k_1\k_2} \pa_{\k_1} \Cb_\r \pa_{\k_2}C^{\r}
+\vf^{\r\s} \pa_{\r} \Cb_\s \pa_{\k}C^{\k}
+\pa_\k\vf^{\r\s} \pa_{\r} \Cb_\s\, C^{\k}
\label{ghshifted}
\nn\\
\eea

\begin{figure}[tbp]
\centering 
\includegraphics[width=.9\textwidth,trim=0 430 0 75,clip]{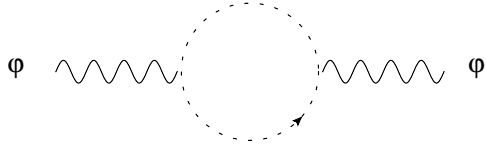}
\caption{background action from ghost loop}
\end{figure}

The background action can be obtained by considering 
\bea
-\fr12<\Big(\int \cL_{\vf CC}\Big)^2>_{1-loop} \la{ghcforBA}
\eea
and integrating out the ghost fields. For example, one of the terms in \rf{ghcforBA} is
\bea
-\fr12<\Big(\int \pa_\k\vf^{\r\s} \pa_{\r} \Cb_\s\, C^{\k}\Big)^2 >_{1-loop}
\label{ctrex}
\eea
and leads to the following background action:
\bea
-\fr{\G(\e)}{(4\pi)^2}\int d^4x\Big(
\fr1{12}\pa_{\s_1}\pa_{\s_2}\f^{\r_1\s_1}\pa_{\r_1}\pa_{\r_2}\f^{\r_2\s_2}
+\fr1{24}\pa_{\s_1}\pa_{\s_2}\f^{\r_1\s_1}\pa^\k\pa_\k \f_{\r_1}^{\s_2}
\Big)  \la{ghBAex}
\eea
where we have switched $\vf$ back to $\f$.
One can straightforwardly show that the divergent part
from \rf{ctrex} is cancelled exactly by the tree-level graph of the background action \rf{ghBAex}.
\begin{figure}[tbp]
\centering 
\includegraphics[width=.7\textwidth,trim=0 450 0 105,clip]{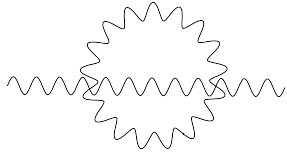}
\caption{one of two-loop diagrams}
\end{figure}

The total one-loop counter term action can be found in \rf{tctr}:
\bea
\D \cL_{1-loop}
&=& \fr{\G(\e)}{(4\pi)^2}\int d^4x\;\Big[\,
\frac{41}{60}\,\pa_\a\pa_\b\f^{\a\b}\,\pa_\g\pa_\d\f^{\g\d}
+\frac{27}{80}\, \pa^\k\pa_\k\f_{\a\b}\,\pa^\l\pa_\l\f^{\a\b}\nn\\
&& -\frac{59}{480}\, \pa^\k\pa_\k\f\,\pa^\l\pa_\l\f+\frac{13}{30}\, \pa^\k\pa_\k \f\,\pa_\g\pa_\d\f^{\g\d}    
  -\frac{27}{40}\,\pa^\k\pa_\k\f_{\a\b}\,\pa^\a\pa^\g\f_{\g}^\b 
\Big]  \nn\\
\eea
where $\e\equiv 2-D/2$.
One can easily check that the counterterms cannot be written in terms of $R$ and $R_{\m\n}$, given that in the leading order,
\bea
R_{\a\b}&=&\fr12 \pa^2 \f_{\a\b}-\fr14 \eta_{\a\b}\pa^2 \f
-\fr12\pa_\a\pa^\g\f_{\b\g}-\fr12\pa_\b\pa^\g\f_{\a\g}\nn\\
R^2&=&\fr14 \pa^2\f\pa^2\f+\pa^2\f \pa_\m\pa_\n\f^{\m\n}
+\pa_\a\pa_\b\f^{\a\b}\pa_\g\pa_\k\f^{\g\k}\nn\\
R^{\a\b}R_{\a\b}&=& \fr14 \pa^2\f_{\a\b}\pa^2\f^{\a\b}  
           -\fr12 \pa^2\f_{\a\b}\pa^\a\pa_\g\f^{\b\g} 
   +\fr12 \pa^2\f \pa_\a\pa_\b\f^{\a\b}  
   +\fr12\pa_\a\pa_\b\f^{\a\b} \pa_\g\pa_\k\f^{\g\k}  \nn\\   
\eea

\subsection{3D analysis}
Our 3D gravity (with two physical degrees of freedom) does not require counterterms at one-loop because Gamma functions can be finite by formal identities in dimensional regularization. However, counterterms are required at two-loop as we will see in the following example.

As in 4D, we consider
\bea
\gt^{mn}\equiv \sqrt{-g}g^{mn}\equiv \eta^{mn}+h^{mn}
\eea
In terms of $h^{mn}$, the quadratic and cubic actions take
\bea
{\cL}_{3D}^\2 &\equiv&   -\fr12\pa_{a}h_{mn}\pa^{a}h^{mn}
+ h^\k h_\k+\fr1{2}\pa_m h\pa^m h \label{3dquadterms}
\eea
\bea
{\cL}_{3D}^\3 &\equiv& -\fr12h^{ab} \pa_a h^{rs}\pa_b h_{rs}
+h_{rs}\pa^a h^{rk}\pa_a h_{k}^s
-h_{rs}\pa_a h^{rb}\pa_b h^{sa} \nn\\
&&\qquad -\fr1{2}\Big(-h^{ab}\pa_a h \pa_b h
+2h_{rs}\pa_a h \pa^a h^{rs}\Big)  \label{3dcubicterms}
\eea
The propagator that follows from \rf{3dquadterms} is
\bea 
<\f_{mn}(y_1)\f_{pq}(y_2)>=\fr12(\eta_{mp}\eta_{nq}+\eta_{mq}\eta_{np}- \eta_{mn}\eta_{pq})  \int \fr{d^3k}{(2\pi)^3}\fr{e^{ik\cdot (y_1-y_2)}}{i k^2}
\eea
 Let us consider an example of the two-loop diagrams drawn in Fig. 4. 
The quartic coupling is given by
\bea
\cL_{3D}^\4&=&\fr12\Big[
(\eta_{k_1k_3}h_{k_2k_4}+\eta_{k_2k_4}h_{k_1k_3})
h^{ab}\pa_ah^{k_1k_2}\pa_bh^{k_3k_4}  \nn\\
&& -(\eta_{k_1k_3}h_{k_2}^s h_{s k_4}
+\eta_{k_2k_4}h_{k_1}^s h_{sk_3}+h_{k_1k_3}h_{k_2k_4})
\eta^{ab}\pa_ah^{k_1k_2}\pa_bh^{k_3k_4}\nn\\
&&+2h_{k_1}^rh_{rk_2}\pa_ah^{k_1b}\pa_bh^{k_2a}
-(\eta_{k_1k_2}h_{k_3k_4}
+\eta_{k_3k_4}h_{k_1k_2})h^{ab}\pa_ah^{k_1k_2}\pa_bh^{k_3k_4}\nn\\
&&+(\eta_{k_1k_2}h_{k_3}^rh_{rk_4}
+\eta_{k_3k_4}h_{k_1}^rh_{rk_2}+h_{k_1k_2}h_{k_3k_4})
\eta^{ab}\pa_ah^{k_1k_2}\pa_bh^{k_3k_4} \Big]
\eea
As in the previous subsections, this can be written as 
\bea
h_{l_1 l_2}h_{l_3 l_4}\pa_{a}h_{ l_5 l_6}\pa_{b}h_{l_7 l_8}
T_{4h}^{a b l_1 l_2 l_3 l_4 l_5 l_6 l_7 l_8}
\eea
where $T_{4h}^{a b l_1 l_2 l_3 l_4 l_5 l_6 l_7 l_8}$ is a numerical tensor whose explicit form can be written easily. The two-loop diagram comes from
the following correlator:
\bea
-\fr12<h_{mn}h_{rs}
\Big(h_{l_1 l_2}h_{l_3 l_4}\pa_{a}h_{l_5 l_6}\pa_{\b}h_{l_7 l_8}
T_{4h}^{a b l_1 l_2 l_3 l_4 l_5 l_6 l_7 l_8}\Big)^2>
\eea
Let us compute
\bea
<h^{mn}h^{rs}\int h^{l_1 l_2}h^{l_3 l_4}\pa^{a}h^{l_5 l_6}\pa^{b}h^{l_7 l_8}
\int h^{l_1' l_2'}h^{l_3' l_4'}\pa^{a'}h^{l_5' l_6'}\pa^{ b'}h^{l_7' l_8'}>
\eea
There are altogether 16 different contractions, some of which are related by simple permutations of indices, that contribute to the two-loop diagrams in Fig. 4. We will illustrate the computations involved with one of the sixteen terms, and pursue the complete evaluation elsewhere.
One of the sixteen terms is given in the momentum space by
\bea
&&\fr{i}2 P^{mnl_1l_2}P^{rs l_1' l_2'}P^{l_3 l_4 l_3' l_4'}
P^{l_5 l_6 l_5' l_6'}P^{l_7 l_8 l_7' l_8'}(2\pi)^4\d(p_1+p_2)
\fr{1}{p_1^2}\fr{1}{p_2^2}\nn\\
&&\int\int \fr{d^3k_3}{(2\pi)^3}\fr{d^3k_5}{(2\pi)^3}
\fr{k_4^a k_4^{a'}k_5^b k_5^{b'}}{k_4^2k_5^2(k_5+k_4-p_1)^2}
\la{2loop}
\eea
Let us evaluate the momentum integrals in two steps; after the Feynman parameterization, the integration over $k_5$ yields
\bea
&&\int \fr{d^3k_5}{(2\pi)^3}
\fr{k_5^b k_5^{b'}}{k_5^2(k_5+k_4-p_1)^2}\nn\\
&=&
\fr{\pi^{\fr32}}{8(4\pi)^{\fr32}}\Big(
3\fr{(k_4^{b_1}-p_1^{b_1})(k_4^{b_2}-p_1^{b_2})}
{[(k_4-p_1)^2]^{1/2}}
-\eta_{b_1 b_2}[(k_4-p_1)^2]^{1/2}\Big)
\eea
Therefore, one gets
\bea
&&\int\int \fr{d^3k_4}{(2\pi)^3}\fr{d^3k_5}{(2\pi)^3}
\fr{k_4^a k_4^{a'}k_5^b k_5^{b'}}{k_4^2k_5^2(k_5+k_4-p_1)^2}\nn\\
&=& \fr{\pi^{\fr32}}{8(4\pi)^{\fr32}}\int \fr{d^3k}{(2\pi)^3}
\Big(
3\fr{k^a k^{a'}(k^{b}-p_1^{b})(k^{b'}-p_1^{b'})}
{k^2[(k-p_1)^2]^{1/2}}
-\eta_{bb'}\fr{k^a k^{a'}[(k-p_1)^2]^{1/2}}{k^2}\Big) 
\label{dmtm} \nn\\
\eea
The integrals in the expression can be evaluated by making repeated use of the following formula \cite{Smirnov}: 
\bea
\int d^Dk \fr{1}{(k^2)^{\xi_1}[(k-p)^2]^{\xi_2}}
=\pi^{D/2}\fr{G(\xi_1,\xi_2)}{(p^2)^{\xi_1+\xi_2-\fr{D}{2}}} \label{smir}
\eea
where $\xi_1,\xi_2$ are arbitrary numbers, and
\bea
G(\xi_1,\xi_2)=\fr{\G(\xi_1+\xi_2-\fr{D}2)\G(\fr{D}{2}-\xi_1)\G(\fr{D}{2}-\xi_2)}{\G(\xi_1)\G(\xi_2)\G(D-\xi_1-\xi_2)}
\eea
where $\G$ represents the Gamma function.
We illustrate the computations involved with the second term in \rf{dmtm}:
\bea
&&\int \fr{d^3k}{(2\pi)^3}\fr{k^a k^{a'}[(k-p_1)^2]^{1/2}}{k^2}\nn\\
&&=
-\fr13\fr{\pa}{\pa (p_1)_{a_1}}\int \fr{d^3k}{(2\pi)^3}\fr{ k^{a'}[(k-p_1)^2]^{3/2}}{k^2}+p_1^{a_1}\int \fr{d^3k}{(2\pi)^3}\fr{ k^{a'}[(k-p_1)^2]^{1/2}}{k^2}  \la{imomi}\nn\\
\eea
Similar steps can be applied to the two terms in \rf{imomi}, and lead to expressions that do not contain any uncontracted indices on the momentum $k$. One can then use the formula \rf{smir}; after some tedious algebra, one gets
\bea
&&\int\int \fr{d^3k_4}{(2\pi)^3}\fr{d^3k_5}{(2\pi)^3}
\fr{k_4^a k_4^{a'}k_5^b k_5^{b'}}{k_4^2k_5^2(k_5+k_4-p_1)^2}\nn\\
&=& \fr{\G(\fr{3-D}{2})}{64\times 105(2\pi)^2}\Big[
(2\eta^{aa'}\eta^{bb'}+\eta^{ab_1}\eta^{a'b'}
+\eta^{ab'}\eta^{a'b})(p_1^2)^2 \nn\\
&&-8(\eta^{aa'}p_1^{b}p_1^{b'}
+\eta^{bb'}p_1^{a}p_1^{a'})(p_1^2)
+4(\eta^{ab'}p_1^{a'}p_1^{b}
+\eta^{a'b'}p_1^{a}p_1^{b}\nn\\
&&+\eta^{ab}p_1^{a'}p_1^{b'}
+\eta^{a'b}p_1^{a}p_1^{b'})(p_1^2)
+8p_1^{a}p_1^{a'}p_1^{b}p_1^{b'}\Big]
\eea
Once this result is substituted into \rf{2loop}, the computation of 
\rf{2loop} is complete.
\section{Discussions}
In this work, we have reviewed and refined the proposal in \cite{Park:2014tia}, where it was noted that the Lagrangian analogues of the Hamiltonian and momentum constraints can be solved. The analysis led to the conclusion that the physical states are three-dimensional at the classical level. The system can be quantized by imposing the canonical commutation relations. In general, quantization introduces fluctuations around the classical vacuum; the fluctuations in the present case were three-dimensional in the operator formalism, and this feature has been carried over to the path integral account of the perturbative analysis that we have developed in this work. The next task was to carry out explicit renormalization in the three-dimensional description. The 3D renormalization at one-loop does not require counterterms in dimensional regularization, thus one will have to consider two-loop renormalization. Most of the computations were done for the 4D case to set the ground for the 3D case; for the three-dimensional two-loop diagrams, we have considered Fig.4 to illustrate the computations involved.

With the help of a Mathematica package xAct`xTensor`, we first carried out one-loop renormalization of the 4D action expanded around a flat spacetime, and compared the results with the those in \cite{Capper:1973pv} obtained long ago. The two results on divergences do not entirely match up; an erroneous Ward-Takahashi type identity used in \cite{Capper:1973pv} should be the reason for the incongruity. We have explicitly obtained the counterterms by applying the background field method to the 4D action expanded around a flat background, and as far as we are aware this is a new result.

Unlike, e.g., \cite{'tHooft:1974bx,Goroff:1985th}, the counterterms that we have obtained are non-covariant and cannot be expressed in terms of covariant quantities such as $R^2$ and/or $R_{\m\n}R^{\m\n}$ expanded around the flat vacuum. 
This then takes us to an issue that is relatively well-understood through a formal argument and explicit examples in non-gravitational theories; namely, the background independence of renormalizability. The fact that the counterterms cannot be expressed in terms of $R^2$ and/or $R_{\m\n}R^{\m\n}$ seems to contradict this expectation. 
The non-covariance of the counterterms was expected from the beginning since the diffeomorphism gets broken once the action is expanded around a flat vacuum.  
What makes the gravitational cases different from non-gravitational cases
is the fact that the counterterms cannot be absorbed into the terms in the expanded action even at one-loop: once the action is expanded around a fixed background, the one-loop renormalizability seems to become more subtle.

As noted in the main body, this unsettling status of the matter should presumably be due to the fact that the counterterms obtained in \cite{'tHooft:1974bx} would formally correspond to the counterterms of the action expanded around the zero metric background.\footnote{Or it could simply be the field redefinition \rf{fred} that caused the problem. This issue is currently under investigation.} With such expansion, explicit perturbative analysis of course cannot be carried out. In this sense, the meaning of the counterterms found in \cite{'tHooft:1974bx} is not entirely clear (at least to us). Nevertheless, the result of \cite{'tHooft:1974bx} does seem to imply that the counterterms for gauge invariant physical quantities would be covariant and come in powers in $R,R_{\m\n}, R_{\m\n\r\s}$. Therefore once the theory is reduced to 3D, renormalizability would be restored in the sense of \cite{'tHooft:1974bx,Goroff:1985th,Anselmi:2003xb}.

The 3D two-loop renormalization would be much more technically demanding than the analysis in the present work. It would be interesting to investigate the possibility that removing the trace piece in the sense of \cite{Park:2014tia} could make things work.  
Of course, the fact that the external legs are not gauge-invariant does not depend on whether or not the trace piece is present. Therefore, even after removing the trace piece in 3D, one cannot a priori expect the counterterms become expressible in terms of $R$ and $R_{\m\n}$. The absence of the trace piece will, however, substantially reduce the number of different types of the counterterms. Also, once the trace piece is removed, all of the unphysical degrees of freedom are gone, therefore, there might be a higher chance that the counterterms will be expressible in terms of $R$ and $R_{\m\n}$. We will report on some of these and/or related issues in the near future.

\newpage
\appendix

\renewcommand{\theequation}{A.\arabic{equation}}
 \setcounter{equation}{0}
\section{notations/conventions and identities}
The signature is mostly plus: 
\bea
\eta_{\m\n}=(-,+,+,+)
\eea
All the Greek indices are four-dimensional
\bea
\a,\b,\g,...,\m,\n,\r...=0,1,2,3
\eea
and all the Latin indices are three-dimensional
\bea
a,b,c,...,m,n,r...=0,1,2
\eea
The perturbative analysis is carried out in terms of the redefined metric:
\bea
\gt^{\m\n}\equiv \sqrt{-g}g^{\m\n}\equiv \eta^{\m\n}+\f^{\m\n}
\la{fred}
\eea
for 4D;
\bea
\tilde{g}^{mn}\equiv \sqrt{-\g}\g^{mn}\equiv \eta^{mn}+h^{mn}
\eea
for 3D.
The following shorthand notations were used:
\bea
\f\equiv \eta_{\m\n}\f^{\m\n}\quad,\quad \f^\m\equiv \pa_\k\f^{\k\m}
\eea
for 4D;
\bea
h\equiv \eta_{mn}h^{mn}\quad,\quad h^m\equiv \pa_kh^{km}
\eea
for 3D. The 4D graviton and ghost propagators are given by
\bea 
<\f_{\m\n}(x_1)\f_{\r\s}(x_2)>&=& P_{\m\n\r\s} \int \fr{d^4k}{(2\pi)^4}\fr{e^{ik\cdot (x_1-x_2)}}{i k^2} \nn\\
<C_\m(x_1)\Cb_\n(x_2)>&=&\eta_{\m\n}\int
 \fr{d^4k}{(2\pi)^4}\;\fr{1}{ik^2}e^{ik\cdot(x_1-x_2)}
\eea
where
\bea
P_{\m\n\r\s}\equiv \fr12(\eta_{\m\r}\eta_{\n\s}+\eta_{\m\s}\eta_{\n\r}- \eta_{\m\n}\eta_{\r\s})
\eea
Similarly, the 3D graviton and ghost propagators are given by
\bea 
<\f_{mn}(y_1)\f_{pq}(y_2)>&=&P_{mnpq}  \int \fr{d^3k}{(2\pi)^3}\fr{e^{ik\cdot (y_1-y_2)}}{i k^2} \nn\\
<C_m(y_1)\Cb_n(y_2)>&=&\eta_{mn}\int
 \fr{d^3k}{(2\pi)^3}\;\fr{1}{ik^2}e^{ik\cdot(y_1-y_2)}
\eea
where
\bea
P_{mnpq}\equiv \fr12(\eta_{mp}\eta_{nq}+\eta_{mq}\eta_{np}- \eta_{mn}\eta_{pq})
\eea
With the Feynman parameterization, one can show 
\bea
&&\int \fr{d^Dl}{(2\pi)^4}\fr{1}{(l-p)^2 l^2} {=} \fr{\G(\e)}{(4\pi)^2}\nn\\
&&\int \fr{d^Dl}{(2\pi)^4}\fr{l_\m}{(l-p)^2 l^2} {=} \fr{\G(\e)}{(4\pi)^2}\fr{p_\m}{2}\nn\\
&&\int \fr{d^Dl}{(2\pi)^4}\fr{l_\m l_\n}{(l-p)^2 l^2} {=} \fr{\G(\e)}{(4\pi)^2}\Big[\fr{1}{3}p_\m p_\n-\fr{1}{12}\d_{\m\n}p^2\Big]\nn\\
&&\int \fr{d^Dl}{(2\pi)^4}\fr{l_\m l_\n l_\r}{(l-p)^2 l^2} {=} \fr{\G(\e)}{(4\pi)^2}\Big[\fr{1}{4}p_\m p_\n p_\r-\fr{1}{24}(\d_{\m\n}p_\r+\d_{\m\r}p_\n+\d_{\n\r}p_\m)p^2\Big]\nn\\
&&\int \fr{d^Dl}{(2\pi)^4}\fr{l_\m l_\n l_\r l_\s}{(l-p)^2 l^2} {=} \fr{\G(\e)}{(4\pi)^2}\Big[\fr{1}{5}p_\m p_\n p_\r p_\s-\fr{1}{40}(\d_{\m\n}p_\r p_\s+\d_{\m\r}p_\n p_\s+\d_{\m\s}p_\n p_\r  \nn\\
 &&+\d_{\n\r}p_\m p_\s+\d_{\n\s}p_\m p_\r+\d_{\r\s}p_\m p_\n)p^2 +\fr{1}{240}(p^2)^2(\d_{\m\n}\d_{\r\s}+\d_{\m\r}\d_{\n\s}+\d_{\m\s}\d_{\r\n})\Big]\nn\\
\eea
where $\e\equiv 2-D/2$.
For the two-loop graph in three dimensions, the following identities were used:
\bea
\int \fr{d^Dk}{(2\pi)^D}\fr{1}{k^2 [(k-p)^2]^{\fr12}}
&=&\fr{\G(\d)}{(2\pi)^{D-1}} \nn\\
\int \fr{d^Dk}{(2\pi)^D}\fr{k^m}{k^2 [(k-p)^2]^{\fr12}}
&=&\fr{\G(\d)}{(2\pi)^{D-1}}\fr{p^m}{3}\nn\\
\int \fr{d^Dk}{(2\pi)^D}\fr{k^m k^n}{k^2 [(k-p)^2]^{\fr12}}
&=&\fr{\G(\d)}{(2\pi)^{D-1}}\Big[
-\fr{1}{15} \eta^{mn}p^2+\fr15 p^m p^n  \Big]\\
\int \fr{d^Dk}{(2\pi)^D}\fr{k^m k^n k^r}{k^2 [(k-p)^2]^{\fr12}}
&=&\fr{\G(\d)}{(2\pi)^{D-1}}\Big[
-\fr{1}{35}( \eta^{mn}p^\r+\eta^{mr}p^n+\eta^{nr}p^m)p^2
+\fr17 p^m p^n p^r \Big] \nn
\eea
\[
\int \fr{d^Dk}{(2\pi)^D}\fr{k^m k^n k^r k^s}{k^2 [(k-p)^2]^{\fr12}}
=\fr{\G(\d)}{(2\pi)^{D-1}}\Big[
\fr{1}{5\times 7\times 9}( \eta^{mn}\eta^{rs}
               +\eta^{mr}\eta^{ns}+\eta^{nr}\eta^{ms})(p^2)^2
               \]
\[               
-\fr1{7\times 9}(\eta^{ms}p^n p^r+\eta^{ns}p^m p^r+
\eta^{rs}p^m p^n+\eta^{mr}p^n p^s+\eta^{mn}p^r p^s+
\eta^{nr}p^m p^s)(p^2)  +\fr19 p^m p^n p^r p^s
\Big]
 \nn\\
\]
where $\d\equiv \fr{3-D}{2}$.

\renewcommand{\theequation}{B.\arabic{equation}}
 \setcounter{equation}{0}
\section{4D case}
Let us shift the field 
\bea
\f^{\m\n}\ra \f^{\m\n}+\vf^{\m\n}
\eea
where $\f$ on the right-hand side denotes the fluctuation 
and $\vf$ the background field. 
\subsection{ghost sector}
Let us put $\cL_{\vf CC}$ in \rf{ghshifted} into two groups:
\bea
\cL_{\vf CC}=\cL_{\vf CC,I}+\cL_{\vf CC,II}
\eea
where
\[
\cL_{\vf CC,I}=-\vf^{\r\k_2} \pa_{\k_1} \Cb_\r \pa_{\k_2}C^{\k_1}
-\vf^{\k_1\k_2} \pa_{\k_1} \Cb_\r \pa_{\k_2}C^{\r}
+\vf^{\r\s} \pa_{\r} \Cb_\s \pa_{\k}C^{\k}
\]
\[
\cL_{\vf CC,II}=\pa_\k\vf^{\r\s} \pa_{\r} \Cb_\s\, C^{\k}
\]
Then $\cL_{\vf CC}^2$ in \rf{ghcforBA} can be grouped into three parts:
\bea
\cL_{\vf CC}^2=\cL_{\vf CC,I}^2+2\cL_{\vf CC,I}\cL_{\vf CC,II}+\cL_{\vf CC,II}^2
\label{Lghsq}
\eea
We have computed the background action from $\cL_{\vf CC,II}^2$ in \rf{ghBAex}: 
\bea
\D\cL^{gh}_{II^2}&=&-\fr{\G(\e)}{(4\pi)^2}\int d^4x\Big(
\fr1{12}\pa_{\s_1}\pa_{\s_2}\f^{\r_1\s_1}\pa_{\r_1}\pa_{\r_2}\f^{\r_2\s_2}
+\fr1{24}\pa_{\s_1}\pa_{\s_2}\f^{\r_1\s_1}\pa^\k\pa_\k \f_{\r_1}^{\s_2}
\Big) \la{ghBAexq} \nn\\ 
\eea
Here we compute the background action resulting from the other two terms in \rf{Lghsq}. Let us first note that $\cL_{\vf CC,I}$ can be rewritten as
\bea
\cL_{\vf CC,I}=\vf_{\l_1\l_2}\pa_{\a_1}\Cb_{\l_3}\pa_{\b_1}C_{\l_4}T_{gh}^{\a_1\b_1\l_1\l_2\l_3\l_4}
\eea
where
\bea
\!\!\!\!T_{gh}^{\a_1\b_1\l_1\l_2\l_3\l_4}\equiv
-\eta^{\l_1\l_3}\eta^{\l_2\b_1}\eta^{\a_1\l_4}
-\eta^{\l_1\a_1}\eta^{\l_2\b_1}\eta^{\l_3\l_4}
+\eta^{\l_1\a_1}\eta^{\l_2\l_3}\eta^{\b_1\l_4}
\eea
The background action from $\cL_{\vf CC,I}^2$ can be obtained by multiplying
$T_{gh}^{\a_1\b_1\l_1\l_2\l_3\l_4}T_{gh}^{\a_2\b_2\l_1'\l_2'\l_3'\l_4'}$ with
the background action that arises from
\bea
-\fr12 \int d^4u d^4v <(\vf_{\l_1\l_2}\pa_{\a_1}\Cb_{\l_3}\pa_{\b_1}C_{\l_4})
(\vf_{\l_1'\l_2'}\pa_{\a_2}\Cb_{\l_3'}\pa_{\b_2}C_{\l_4'})>;
\eea
a straightforward calculation leads to the following background action
\bea
&&-\fr12 \fr{\G(\e)}{(4\pi)^2}\int d^4x\; \eta_{\l_3\l_4'}\eta_{\l_4\l_3'}
\Big[\;\fr1{30}\pa_{\a_1}\pa_{\b_1}\f_{\l_1\l_2}\pa_{\a_2}\pa_{\b_2}\f_{\l_1'\l_2'} \nn\\
&&+\fr1{60} \pa_\k\pa^\k \f_{\l_1\l_2}(\eta_{\a_1\a_2}\pa_{\b_1}\pa_{\b_2}+
\eta_{\a_1\b_1}\pa_{\a_2}\pa_{\b_2}+\eta_{\a_2\b_2}\pa_{\a_1}\pa_{\b_1}
+\eta_{\b_1\b_2}\pa_{\a_1}\pa_{\a_2})\f_{\l_1'\l_2'}\nn\\
&&-\fr{1}{40}\pa_\k\pa^\k \f_{\l_1\l_2}(\eta_{\a_1\b_2}\pa_{\a_2}\pa_{\b_1}
+\eta_{\a_2\b_1}\pa_{\a_1}\pa_{\b_2})\f_{\l_1'\l_2'}\nn\\
&&+\fr{1}{240}(\eta_{\a_1\b_2}\eta_{\a_2\b_1}+\eta_{\a_1\b_1}\eta_{\a_2\b_2}
+\eta_{\a_1\a_2}\eta_{\b_1\b_2})\pa_\k\pa^\k \f_{\l_1\l_2}\pa_\d\pa^\d \f_{\l_1'\l_2'}
\Big]  \la{dLgh1}
\eea
Taking the tensors $T_{gh}^{\a_1\b_1\l_1\l_2\l_3\l_4}T_{gh}^{\a_2\b_2\l_1'\l_2'\l_3'\l_4'}$ into account, one gets
\bea
\D\cL^{gh}_{I^2}&=&-\frac{1}{60}\, \pa^\g\pa_\g\f_{\a\b}\,\pa^\k\pa_\k\f^{\a\b}-\frac{1}{120} \, \pa^\g\pa_\g \f \, \pa^\k\pa_\k\f
-\frac{1}{15}\,\pa_\a\pa_\b \f^{\a\b}\,\pa_\g\pa_\k\f^{\g\k}\nn\\
 &&-\frac{1}{20}\,\eta^{\b\k} \pa^\g\pa_\g \f_{\a\b}\pa^\a\pa^\d\f_{\d\k}   
 +\frac{1}{60}\,\pa^\g\pa_\g \f\, \pa_\a\pa_\b\f^{\a\b}
 \label{ghBApart2}
\eea
The background action from $2\cL_{\vf CC,I}\cL_{\vf CC,II}$ sector is
\bea
\D\cL^{gh}_{I\,II}=\frac{1}{12}\, \pa^r\pa_r \f_{mn}\, \pa^m\pa^p\f_{pq}\eta^{nq}-\frac{1}{8}\,  \pa^r\pa_r\f\, \pa_m\pa_n\f^{mn}-\frac{1}{12}\, \pa_m\pa_n\f^{mn}\, \pa_p\pa_q\f^{pq} \nn\\\label{ghBApart3}
\eea
The total background action from the ghost sector is sum of \rf{ghBAexq}, \rf{ghBApart2} and \rf{ghBApart3}:
\bea
\D\cL^{gh}=\D\cL^{gh}_{I^2}+\D\cL^{gh}_{I\,II}+\D\cL^{gh}_{II^2}
\eea
 They will be added to the background action arising from the graviton sector, which we will now examine.

\subsection{graviton sector}
Upon substituting this into the cubic terms and collecting the terms that contain one factor of $\vf$, one gets
\bea
\cL_{\vf\f\f}&=& -\pa_\a \vf^{\r\s}\pa_\b \f_{\r\s}\f^{\a\b}+2\pa^\a \vf^{\r\k}\pa_\a\f_\k^\s\f_{\r\s}-2\pa_\b \vf^{\s\a}\pa_\a\f^{\r\b}\f_{\r\s}\nn\\
&&+\fr12 \pa_\a\vf \pa_\b\f \f^{\a\b}
-\fr12 \pa^\a \vf^{\r\s}\pa_\a\f\, \f_{\r\s}
-\fr12 \pa_\a \vf\pa^\a\f^{\r\s}\f_{\r\s}\nn\\
&&-\fr12 \vf^{\a\b}\pa_\a\f^{\r\s}\pa_\b \f_{\r\s}
+\vf_{\r\s}\pa^\a\f^{\r\k}\pa_\a \f_\k^\s
-\vf_{\r\s}\pa_\a\f^{\r\b}\pa_\b \f^{\s\a}\nn\\
&&+\fr14\vf^{\a\b}\pa_\a\f\pa_\b \f
-\fr12\vf_{\r\s}\pa_\a\f\pa^\a\f^{\r\s}
\eea
The background action from this is given by
\bea
\D \cL_{grav}&=&-\fr12 \fr{\G(\e)}{(4\pi)^2} \Big[-\frac{11}{6}\,\pa_m\pa_n\f^{mn}\,\pa_p\pa_q\f^{pq}-\frac{17}{24}\, \pa^r\pa_r\f_{mn}\,\pa^s\pa_s\f^{mn}\nn\\
&& +\frac{11}{48}\, \pa^r\pa_r\f\,\pa^s\pa_s\f-\frac{13}{12}\, \pa^r\pa_r \f\,\pa_p\pa_q\f^{pq}    
  +\frac{4}{3}\eta^{nq}  \pa^r\pa_r\f_{mn}\,\pa^m\pa^p\f_{pq} \Big]\nn\\
\eea
\subsection{total background action}
Combining the ghost and graviton sectors, one gets for the total one-loop
\bea
\D \cL_{1-loop}&=& \D \cL_{grav} +\D\cL^{gh}_{I^2}+\D\cL^{gh}_{I\,II}+\D\cL^{gh}_{II^2} \nn\\
&=& \fr{\G(\e)}{(4\pi)^2}\int d^4x\Big(-\fr12\Big) \Big[-\frac{11}{6}\,\pa_m\pa_n\f^{mn}\,\pa_p\pa_q\f^{pq}-\frac{17}{24}\, \pa^r\pa_r\f_{mn}\,\pa^s\pa_s\f^{mn}\nn\\
&& +\frac{11}{48}\, \pa^r\pa_r\f\,\pa^s\pa_s\f-\frac{13}{12}\, \pa^r\pa_r \f\,\pa_p\pa_q\f^{pq}    
  +\frac{4}{3}\eta^{nq}  \pa^r\pa_r\f_{mn}\,\pa^m\pa^p\f_{pq} \Big]\nn\\
&&-\Big(
\fr1{12}\pa_{s_1}\pa_{s_2}\f^{r_1s_1}\pa_{r_1}\pa_{r_2}\f^{r_2s_2}
+\fr1{24}\pa_{s_1}\pa_{s_2}\f^{r_1s_1}\pa^k\pa_k \f_{r_1}^{s_2}
\Big)  \nn\\
&&+\Big(-\frac{1}{60}\, \pa^r\pa_r\f_{mn}\,\pa^s\pa_s\f^{mn}-\frac{1}{120} \, \pa^r\pa_r \f \, \pa^s\pa_s\f
-\frac{1}{15}\,\pa_m\pa_n \f^{mn}\,\pa_p\pa_q\f^{pq}\nn\\
 &&\hspace{1.2in}-\frac{1}{20}\,\eta^{nq} \pa^r\pa_r \f_{mn}\pa^m\pa^p\f_{pq}   
 +\frac{1}{60}\,\pa^r\pa_r \f\, \pa_m\pa_n\f^{mn}\Big)\nn\\
&&+\Big(\frac{1}{12}\, \pa^r\pa_r \f_{mn}\, \pa^m\pa^p\f_{pq}\eta^{nq}-\frac{1}{8}\,  \pa^r\pa_r\f\, \pa_m\pa_n\f^{mn}-\frac{1}{12}\, \pa_m\pa_n\f^{mn}\, \pa_p\pa_q\f^{pq}  \Big)  \nn\\
&=& \fr{\G(\e)}{(4\pi)^2}\int d^4x\;\Big[\,
\frac{41}{60}\,\pa_\a\pa_\b\f^{\a\b}\,\pa_\g\pa_\d\f^{\g\d}
+\frac{27}{80}\, \pa^\k\pa_\k\f_{\a\b}\,\pa^\l\pa_\l\f^{\a\b}\nn\\
&& -\frac{59}{480}\, \pa^\k\pa_\k\f\,\pa^\l\pa_\l\f+\frac{13}{30}\, \pa^\k\pa_\k \f\,\pa_\g\pa_\d\f^{\g\d}    
  -\frac{27}{40}\,\pa^\k\pa_\k\f_{\a\b}\,\pa^\a\pa^\g\f_{\g}^\b 
\Big]   \la{tctr}
\eea

\newpage

\end{document}